\newcommand{\PaperNumber}{XXX}

\documentclass[screen, nonacm]{acmart} % FOR ARXIV
\makeatletter
\def\@ACM@checkaffil{% Only warnings
    \if@ACM@instpresent\else
    \ClassWarningNoLine{\@classname}{No institution present for an affiliation}%
    \fi
    \if@ACM@citypresent\else
    \ClassWarningNoLine{\@classname}{No city present for an affiliation}%
    \fi
    \if@ACM@countrypresent\else
        \ClassWarningNoLine{\@classname}{No country present for an affiliation}%
    \fi
}
\makeatother
\setcopyright{none}

\usepackage{microtype}
\usepackage{graphicx}
\usepackage{subfigure}
\usepackage{booktabs} 
\usepackage{multirow}

\usepackage{hyperref}
\usepackage{caption}
\usepackage{xcolor}
\usepackage{color}

\usepackage{amsmath,amscd,url,enumerate, mathtools}
\usepackage[utf8]{inputenc}
\usepackage{wrapfig}
\usepackage{pdflscape}
\usepackage{tikz}
\usetikzlibrary{decorations.markings,arrows}
\usetikzlibrary{positioning,arrows,fit}

\usepackage{tikz-cd}
\usepackage{amsfonts}
\usepackage{graphicx}
\usepackage{mathrsfs}
\usepackage{dsfont}
\usepackage{enumitem}
\usepackage{array}
\usepackage{verbatim}
\newcolumntype{?}{!{\vrule width 1pt}}
\usepackage[capitalize,noabbrev]{cleveref}
\usepackage[font=footnotesize,labelfont=bf, labelsep=period, textfont=sl]{caption}
\usepackage{pythonhighlight} % To put Python code in Appendix.

\newcommand{\para}[1]{{\vspace{2pt} \bf \noindent #1}}  
\newenvironment{packed_itemize}{
\begin{list}{\labelitemi}{\leftmargin=1em}
\setlength{\itemsep}{1pt}                                                           
\setlength{\parskip}{0pt}                                                                                 \setlength{\parsep}{0pt}                                                                                  \setlength{\headsep}{0pt}                                                                                 \setlength{\topskip}{0pt}                                                                                 \setlength{\topmargin}{0pt}                                                                               \setlength{\topsep}{0pt}                                                                                  \setlength{\partopsep}{0pt}                                                                               }{\end{list}}

\newcommand{\orig}[1]{$\mathcal{O}_{\text{#1}}$}
\newcommand{\origm}[1]{\mathcal{O}_{\text{#1}}}

\title[Creative Homogeneity Across LLMs]{We're Different, We're the Same: 
 Creative Homogeneity Across LLMs}
 \author{Emily Wenger}
 \authornote{Correspondence to: emily.wenger@duke.edu}
 \affiliation{
    \institution{Duke University}
}
\author{Yoed Kenett}
\affiliation{
    \institution{Technion - Israel Institute of Technology}
}

\makeatletter
\let\@authorsaddresses\@empty
\makeatother

\begin{abstract}

Numerous powerful large language models (LLMs) are now available for use as writing support tools, idea generators, and beyond. Although these LLMs are marketed as helpful creative assistants, several works have shown that using an LLM as a creative partner results in a narrower set of creative outputs. However, these studies only consider the effects of interacting with a single LLM, begging the question of whether such narrowed creativity stems from using a particular LLM\textemdash which arguably has a limited range of outputs\textemdash or from using LLMs {\em in general} as creative assistants. To study this question, we elicit creative responses from humans and a broad set of LLMs using standardized creativity tests and compare the population-level diversity of responses. We find that LLM responses are much more similar to other LLM responses than human responses are to each other, even after controlling for response structure and other key variables. This finding of significant homogeneity in creative outputs across the LLMs we evaluate adds a new dimension to the ongoing conversation about creativity and LLMs. If today's LLMs behave similarly, using them as a creative partners\textemdash regardless of the model used\textemdash may drive all users towards a limited set of ``creative'' outputs. 

\end{abstract}

\begin{document}

\maketitle

\section{Introduction}
\label{sec:intro}

Large language models (LLMs) have moved out of research labs and into our everyday lives. Given their advanced abilities to generate text and respond to prompts, LLMs are often marketed as creativity support tools that allow users to write drafts, edit documents, and generate novel ideas with ease~\cite{gemini_ad, apple_ad, notion_ad, grammarly_ad}. Consumers have responded eagerly to these suggestions. According to a 2024 survey by Adobe, over half of Americans have used generative AI tools like LLMs as creative partners for brainstorming, drafting written content, creating images, or writing code. An overwhelming majority of LLM users surveyed believe these models will help them be more creative~\cite{adobe_survey}. 

% Wrinkle: recent work shows that using an LLM as a creativity support tool can homogenize content. 
While appealing, outsourcing our creative thinking to LLMs could have unintended consequences and demands further scrutiny. For example, recent work has unearthed complications around the use of LLMs as creativity support tools. Researchers found that LLM-aided creative outputs look individually creative but are often quite similar to other LLM-aided outputs. Such ``homogeneity'' in LLM-aided creative outputs has been observed in a variety of settings, from creative writing to online survey responses to research idea generation and beyond~\cite{anderson2024homogenization, si2024can, zhang2024generative, Doshi_Hauser_2024, moon2024homogenizing}. 

While concerning, these works typically only look at a single LLM and it's effect on downstream creative content. In a prototypical example, ~\citet{Doshi_Hauser_2024} compared the individual and collective creativity of two groups of writers\textemdash humans alone and humans aided by ChatGPT\textemdash and found that stories produced by the ChatGPT-aided group were more homogeneous. Related work from Moon, Green, and Kushlev~\cite{moon2024homogenizing} compared college essays written by humans and GPT models and found that LLM-authored essays contributed fewer new ideas and were more homogeneous than human-authored essays. However, such work begs the question: does the observed homogeneity occur because only a single type of LLM (GPT variants) is studied? It could be reasonably argued that a single LLM must have a limited range of outputs, causing the homogeneity. Perhaps if writers all used different LLMs, creativity would be restored. %, because each model gives different responses.} 

% Or, is it the case that us? Our paper asks and begins to answer the latter question.
Recent work studying feature space alignment in LLMs suggests otherwise. There is a long line of work measuring feature space similarity in machine learning models, since this is believed to indicate overall model similarity~\cite{ sucholutsky2023getting, kornblith2019similarity, lenc2015understanding, Liang_Li_Li_Wang_Zhang_2020, bansal2021revisiting}. Some initial work has applied these techniques to large-scale LLMs and found evidence of ``feature universality'' in these models~\cite{lan2024sparse, klabunde2023towards, huh2024platonic}. 
We postulate that such feature space alignment in LLMs may result in homogeneous creative outputs {\em across} these models. This would imply that the use of LLMs as creative partners {\em in general} leads to output homogeneity, because all LLMs would have limited and similar output ranges. 

The consequences of cross-LLM homogeneity would be significant in the creative space and beyond. Humans who rely on LLMs as creative partners would find their creative outputs remarkably similar to those of other LLM users regardless of the model used, resulting in a collective narrowing of societal creativity. More broadly, homogeneity among widely used LLMs could lead to bias propagation, widespread security vulnerabilities, or other problems~\cite{kleinberg2021algorithmic, bommasani2022picking}. 

\para{Our Contribution.} This work explores possible convergence in the creative outputs of large-scale LLMs. We test this by soliciting creative outputs from LLMs and humans using standardized creativity tests\textemdash the Alternative Uses Task~\cite{guilford1978alternate}, Forward Flow~\cite{gray2019forward}, and the Divergent Association Task~\cite{olson2021naming}\textemdash and measuring the population-level variability of responses. While caution should be used in extrapolating human-centric psychological tests to non-human entities (see \S\ref{sec:method}), these tests are useful in our setting because of their standardized output format. This allows us to disambiguate similarity in response structure from similarity in response content, the true goal. Our analysis shows that:
\vspace{-0.1cm}
\begin{itemize}
\item Mirroring prior work~\cite{hubert2024current}, {\em LLMs match or outperform humans on standard tests of individual creativity}.
\item Yet, this finding of individual creativity is misleading because {\em LLM responses to creative prompts are much more similar to each other than are human responses}, even after controlling for LLM ``family'' overlap and differences in human/LLM response structure.
\item {\em Altering the LLM system prompt} to encourage higher creativity {\em slightly increases overall LLM creativity and inter-LLM response variability}, but human responses are still more variable. 
\end{itemize}

\para{Implications.} We believe these findings highlight a potential danger of relying on generative AI models as creative partners. If today's most popular models exhibit a high degree of overlap in creative outputs, using any of them to aid creativity\textemdash as will happen if these models are integrated into platforms we regularly use for writing or creative thinking\textemdash could self-limit us from reaching the divergent creativity that defined artistic geniuses like Mozart, Shakespeare, and Picasso. Our set of AI ``creative'' partners will instead collectively drive us towards a mean.

\section{Related Work}
\label{sec:related}

\para{Creativity, Homogeneity, and LLMs.} Prior work has explored issues of creativity and homogeneity related to specific LLMs. Several works have compared human and LLM performance on standard creativity tests, typically using GPT models, and found that LLMs often outperform humans on these tests~\cite{hubert2024current, Chen_Ding_2023, stevenson2022putting}. Despite LLMs' displays of individual creativity, numerous studies have shown that using LLMs to support creative tasks tends to homogenize creative outputs. For example,~\citet{Doshi_Hauser_2024} found that writers who used GPT-4 as a creativity support tool produced more creative stories than humans working alone, but the stories from writers who collaborated with GPT-4 were more similar to each other than were stories from human writers. This phenomenon of LLM-drive content homogenization appears across domains\textemdash in research idea generation~\cite{si2024can}, essay writing~\cite{moon2024homogenizing}, survey responses~\cite{zhang2024generative}, creative ideation~\cite{anderson2024homogenization}, and art~\cite{zhou2024generative}. Recent work also showed that when GPT models are evaluated multiple times on creativity tests like the DAT, their responses tend to overlap, even if each individual response achieves a high ``creativity'' score~\cite{cropley2023artificial}. Such findings further motivate our study of whether it is the use of specific models in these studies\textemdash often ChatGPT\textemdash that causes observed homogeneity, or if such homogeneity would be observed {\em regardless of the model used}. 

Finally, a few works have considered issues of monoculture related to machine learning algorithms. Several works demonstrate suboptimal outcomes when multiple firms employ the same algorithm for decision-making~\cite{kleinberg2021algorithmic,bommasani2022picking}.~\cite{wu2024generative} proposed the term ``generative monoculture'' to describe the narrow distribution of LLM outputs relative to that of their training data\textemdash an observation related to the creative narrowing observed in other work. However, none of these works considered similarity {\em across} models. %  ~\cite{zhou2024shared} put two LLMs in conversation about fake ideas and found that they respond earnestly to the other's hallucinations, and from this conclude that LLMs share a hallucination space.

\para{LLM Similarity.} Numerous papers have worked to measure similarity between model feature representations, primarily in classifiers~\cite{sucholutsky2023getting, kornblith2019similarity, lenc2015understanding, Liang_Li_Li_Wang_Zhang_2020, bansal2021revisiting}. Such similarity is believed to indicate overall similarity between models and could lead to interesting downstream consequences, such as attack vectors that transfer between models (e.g.~\cite{szegedy2013intriguing, wang2018great, papernot2016transferability, demontis2019adversarial} among many others). Nascent work applies similar methods to LLMs and finds evidence of ``feature universality'' across LLMs~\cite{lan2024sparse, klabunde2023towards}.~\citet{huh2024platonic} also measured feature space alignment between open-source LLMs and postulated that large models will inevitably become more similar over time. However, limited work has considered downstream consequences of LLM similarity. One work~\cite{jeong2024bias} examines question-answering bias of 10 LLMs across 4 ``families'', but finds little evidence of bias similarity among models. One paper demonstrates jailbreak attacks that transfer between LLMs~\cite{zou2023universal} but does not specifically leverage LLM similarity in attack development. %To our knowledge, one has  root cause of this feature space similarity, particularly the dataset overlap. demonstrate the possibility of ``model stitching,'' where two deep neural networks trained on similar datasets are combined via a single trainable layer. This stitching is possible because the models have similar feature representations spaces. More recently,~\cite{huh2024platonic} proposed a ``Platonic Representation Hypothesis,'' which states that as models grow larger and are trained on more data, their knowledge and internal feature representations will become more similar.  

\para{This paper.} We build on this prior work to study {\em creative output variability across LLMs}. Several works have shown that using {\em specific} LLMs as creative partners narrows the range of creative outputs~\cite{Doshi_Hauser_2024, medieroshuman, anderson2024homogenization,cropley2023artificial,moon2024homogenizing}. We instead evaluate the diversity of responses to creative prompts across {\em many} LLMs using standard creativity tests and compare this to the diversity of human responses. We believe this study will enhance the current debate surrounding LLMs and creativity, clarifying whether it is the use of a specific LLM that homogenizes creative outputs or the use of LLMs in general. 
\section{Methodology}
\label{sec:method}

 Our goal is to measure whether LLMs produce more, less, or equally diverse creative outputs as a group of humans. We measure this diversity (or variability) in responses by computing the semantic similarity among responses of humans and LLMs to prompts designed to elicit creativity.  This section describes the creativity prompts we use, humans and LLMs tested, and evaluation metrics. 

\subsection{How do we elicit creative responses from LLMs?}
\label{subsec:creative}

The American Psychological Association defines creativity as ``the ability to produce or develop original work, theories, techniques, or thoughts''~\cite{apa-homepage}. Since our goal is to compare the diversity of creative responses from LLMs and humans, we sought out methods to elicit and compare creative outputs. Given the novelty of this field, no standard benchmarks exist for comparing LLM and human creativity. However, prior work has applied tests of divergent thinking in humans, which elicit qualities psychologists view as important to creativity, to LLMs and found that LLMs like ChatGPT scored similarly to humans~\cite{stevenson2022putting, Chen_Ding_2023, hubert2024current, zhao2024assessing}.

\para{Creativity tests for humans.} One of the original divergent thinking tests was Guilford's Alternative Uses Test (AUT)~\cite{guilford1978alternate}, which presents subjects with an object and asks them to describe creative uses for it. AUT responses are scored by measuring the number of different uses presented (``fluency''), the originality of those ideas (``originality''), how different they are from each other (``flexibility''), and the level of detail provided (``elaboration''). While effective, the AUT evaluation process is onerous, so researchers have developed more lightweight divergent thinking tests in recent years. One popular test, Forward Flow~\cite{gray2019forward} (FF), measures the divergence of a user's chain of thought from a fixed starting point. Another, the Divergent Association Test (DAT)~\cite{olson2021naming}, asks subjects to list 10 unrelated words. Both capture similar characteristics to the AUT but with less burden on participants and evaluators. 

\para{Should we run human creativity tests on LLMs?} Given our goals, it seems reasonable to test humans and LLMs using the AUT, FF, and DAT and then compare the population-level variability of their responses. However, it is an active area of debate as to whether psychological tests\textemdash including those of creativity\textemdash designed for humans are applicable to LLMs. Some argue that works probing LLM performance on these tests is misguided due to fundamental differences between LLM and humans~\cite{stella2023using, gupta2024self, cropley2023artificial}. Future scholarship will inevitably continue this debate. 

\para{Our approach.} Despite this disagreement, we believe creativity tests administered to LLMs can be useful for our purposes, {\em because we are not trying to measure inherent LLM creativity.} Instead, we would like to understand the variability of LLM and human responses to creative prompts, which is an empirical rather than psychological trait. Well-trained LLMs should respond similarly to factual questions but not necessarily to creative prompts, so if there is indeed homogeneity in LLM outputs, creativity is the appropriate lens through which to evaluate this question.

However, the question remains of what type of creative prompts to use. An obvious approach is asking LLMs to write short stories (similar to~\cite{Doshi_Hauser_2024}) and comparing the variability of these to that of human-composed stories, but this approach has a notable caveat. We need to disentangle similarity in the {\em structure} of creative responses from similarity in their {\em content}\textemdash the latter matters significantly, but former does not. If LLMs share certain output quirks in generated text, like using the passive tense or gerunds, we do not want these to skew measurements of LLM response variability. Creativity tests like the AUT, DAT, and FF provide a helpful solution to this potential problem\textemdash they are designed to elicit creative outputs in a structured manner. Therefore, we use these tests in our analysis, but future work should consider other ways to elicit easily comparable creative responses from humans and LLMs. 

\subsection{Tests We Use}
\label{subsec:tests}

Based on the reasoning of \S\ref{subsec:creative}, we compare the variability of human and LLM responses to the AUT, FF, and DAT tests. Exact test wording is in Appendix~\ref{appx:dat_prompts}. 

\para{Guilford's Alternative Uses Test (AUT)~\cite{guilford1978alternate}} presents people with an object and asks them to write down as many creative uses for it as they can think of. Following established best practices~\cite{dumas2014understanding, barbot2018dynamics}, we test users with five common objects\textemdash book, fork, table, hammer, and pants. Using multiple starting objects reduces the effect of a particular object (e.g. book) on participant responses, ensuring results generalize~\cite{dumas2021measuring}. It also allows us to collect more data, given the limited number of LLMs we can evaluate relative to the number of possible human subjects. 

\para{Forward Flow~\cite{gray2019forward}} measures how much a person's thoughts diverge from a given starting point. It provides a starting word and asks people to write down the next word that follows in their mind from the previous word for up to 20 words. We follow the original Forward Flow paper and run our study using five different start words: candle, table, bear, snow, and toaster. As in the AUT, providing multiple creative stimuli ensures results generalize and gives us more data. 

\para{The Divergent Association Task (DAT)~\cite{olson2021naming}} asks subjects to list 10 words that are as unrelated as possible. These are subject to certain constraints: only nouns, no proper nouns, only single words in English, and the task must be completed in less than four minutes. The DAT provides a limited amount of information compared to the other tests, since the creative stimulus cannot be varied.

\subsection{Test subjects}  
We administer these tests to a set of LLMs and a set of humans, following IRB-approved user study protocol.

\para{Large Language Models.} As a baseline, we test 22 large language models with public APIs\footnote{https://docs.github.com/en/github-models/prototyping-with-ai-models}: {\em AI21-Jamba-Instruct}, {\em Cohere Command R}, {\em Cohere Command R Plus}, {\em Meta Llama 3 70B Instruct}, {\em Meta Llama 3 8B Instruct}, {\em Meta Llama 3.1 405B Instruct}, {\em Meta Llama 31 70B Instruct}, {\em Meta Llama 3.1 8B Instruct}, {\em Mistral large}, {\em Mistral large 2407}, {\em Mistral Nemo}, {\em Mistral small}, {\em Google Gemini 1.5}, {\em gpt 4o}, {\em gpt 4o mini}, {\em Phi 3 medium 128k instruct}, {\em  Phi 3 medium 4k instruct}, {\em Phi 3 mini 128k instruct}, {\em Phi 3 mini 4k instruct}, {\em Phi 3 small 128k instruct},{\em  Phi 3 small 8k instruct,} and {\em Phi 3.5 mini instruct}.  Models in the same ``family'' (e.g. all Llamas, all GPTs, etc) may generate unusually similar responses due to similarities in architecture, training data, or optimization techniques. To control for this, we restrict ourselves following subset of models when conducting statistical tests, which contains only one model from each ``family'': {\em AI21 Jamba 1.5 Large}~\cite{team2024jamba}, {\em Google Gemini 1.5}~\cite{team2024gemini}, {\em Cohere Command R Plus}~\cite{cohere}, {\em Meta Llama 3 70B Instruct}~\cite{dubey2024llama}, {\em Mistral Large}~\cite{jiang2023mistral}, {\em gpt 4o}~\cite{achiam2023gpt}, and {\em Phi 3 medium 128k Instruct}~\cite{abdin2024phi}. All these models were trained by distinct entities, providing a reasonable independent baseline. In \S\ref{sec:ablate}, we also explore how models with the same ``family'' behave. For these experiments, we use the Llama model family~\cite{dubey2024llama}: {\em Meta Llama 3 70B Instruct}, {\em Meta Llama 3 8B Instruct}, {\em Meta Llama 3.1 405B Instruct}, {\em Meta Llama 31 70B Instruct}, {\em Meta Llama 3.1 8B Instruct}. We evaluate all models with the default system prompt of ``You are a helpful assistant'' but explore the effect of varying this in \S\ref{sec:ablate}. % \todo{We call the base set X and the subset of non-overlapping models Y}
After obtaining model responses, we remove unnecessary punctuation (e.g. numbered DAT outputs).

\para{Human subjects.} We use two sources of human responses as a ground truth set for human creativity. First, we run an IRB-approved user study\footnote{IRB information redacted for anonymous submission}. Study subjects were recruited from the Prolific platform were asked to complete the DAT, FF, and AUT creativity tests (see Appendix~\ref{appx:dat_prompts} for study wording). It took participants 19 minutes on average to
complete the survey, and they were compensated at a rate of \$15/hour. Participant demographics are described in Table~\ref{tab:demo}. All patients completed a consent form before starting the study. We recruited $114$ initial participants from the Prolific platform, screening for English fluency and an approval rating $>95$. Of these, $12$ were removed on suspicion of being bots due to unusually short response times ($<5$ minutes) or failed attention checks, so the final dataset contains $102$ human responses. Authors manually inspected responses to correct obvious misspellings.  
\begin{table}[h]
\centering
\begin{tabular}{ll@{\hskip 0.3in}ll@{\hskip 0.3in}ll}
\toprule
\multicolumn{2}{c}{\hspace{-0.3in} \bf Age} & \multicolumn{2}{c}{\hspace{-0.3in} \bf Gender} & \multicolumn{2}{c}{\hspace{-0.3in} \bf Race}                                \\ \midrule
18-24       & 22\%      & Female          & 51\%     & Asian                                            & 6\%  \\
25-34       & 31\%      & Male            & 46\%     & Black or African American                        & 28\% \\
35-44       & 23\%      & Non-Binary      & 3\%      & Hispanic or Latino or Spanish Origin of any race & 11\% \\
45-54       & 19\%      &                 &          & White                                            & 53\% \\
55+         & 5\%       &                 &          & Other                                            & 2\% \\ \bottomrule
\end{tabular}%
%}
\vspace{0.1cm}
\caption{Demographics of human study participants.}
\label{tab:demo}
\vspace{-0.3cm}
\end{table}

The risk in relying on responses from online crowd workers is that they may themselves be bots or may leverage LLMs in crafting their responses, resulting in reduced response diversity~\cite{zhang2024generative}. Prolific runs strict tests to ensure human responses, and we also used safeguards to prevent this, including attention tests and post-hoc data inspection. However, the risk remains. Therefore, we use public datasets of human responses to the AUT~\footnote{https://osf.io/u3yv4}, FF~\footnote{https://osf.io/7p5mt/}, and DAT~\footnote{https://osf.io/kbeq6/} from prior work as a secondary validation dataset. These data are from creativity tests run in person before the rise of LLMs (around 2022), so they are unlikely to contain LLM responses. However, given their public nature, these datasets may have been used to train LLMs, resulting in unusual similarity between LLM and these human responses. We use data collected in our user study for our main analysis in \S\ref{sec:results} but re-run population-level originality tests with this data in \S\ref{sec:ablate} for validation. 

\subsection{Evaluation Metrics}
\label{sec:metrics}

The primary goal of this study is to evaluate the {\em variability} of LLM responses to creative prompts relative to that of humans. To do this, we compute the semantic similarity of responses in different populations (LLMs vs. humans) and compute distributional differences in similarity scores between populations. As a baseline, we also compare the {\em originality} of individual LLM responses to the tests relative to that of humans. 

\subsubsection{Scoring individual originality.}
\label{sec:indiv_orig}

Although divergent thinking tests can be measured using multiple metrics, it has long been argued the {\em originality} of responses is the strongest indicator of creativity~\cite{mednick1962associative}. 
Originality\textemdash how novel tests responses are relative to the given prompt(s)\textemdash can also easily be measured in an automated fashion by embedding prompts and responses in a mathematical feature space and measuring the cosine distance between the feature vectors~\cite{beaty2014roles}. Prior work confirms that such automated analysis closely matches originality rankings of human scorers~\cite{dumas2021measuring}.

Our metrics for individual originality follow the guidelines of the original studies but use the automated evaluation methods of~\cite{dumas2021measuring}, including the use of the GloVe 840B model~\cite{pennington2014glove} to compute word embeddings. 
The format of each test necessitates different originality scoring procedures, described in detail in Appendix \S\ref{appx:originality}. Originality scores are denoted as \orig{t}$(\mathcal{P})$, where $t$ = AUT, FF, or DAT and $\mathcal{P}$ is a population, either humans or LLMs. 

\para{Distributional differences.} After computing originality scores, we can then compare the distributions of $\mathcal{O}_t(LLM)$ and $\mathcal{O}_t(Humans)$ to measure differences in originality between the two groups.  We do this by testing for statistically significant differences in $\mu(\mathcal{O}_t(LLM))$ and  $\mu(\mathcal{O}_t(Human))$ using Welch's t-test to compare differences in means, since the populations typically do not exhibit equal variance. We use a statistical significance threshold of $\rho=0.01$. For all tests, the null hypothesis is that $\mu(\mathcal{O}_t(LLM)) = \mu(\mathcal{O}_t(Human))$, and the alternative is that  $\mu(\mathcal{O}_t(LLM)) > \mu(\mathcal{O}_t(Human))$.

\subsubsection{Scoring population-level variability.}  We measure the variability in responses to the creativity tests from a given population %this two ways: 
by computing the semantic distances between sets of responses from individuals in the population (e.g. comparing the set of AUT uses produced by an LLM to that of another LLM). 
If many population members given semantically similar sets of answers, this indicates that the response variability of the population is low, and vice versa if it is high. We denote the variability of a population $\mathcal{P}$ on test $t$ as $\mathcal{V}_t(\mathcal{P})$, the set of all similarity scores between all responses from all population members. As before, $\mathcal{P}$ refers to either LLMs or humans. 

We use a sentence embedding model $\mathcal{S}$ to measure semantic similarity between responses. Sentence embedding models map sentences or short paragraphs to feature vectors and, similar to the word embedding model, map similar content to similar feature vectors. We compute elements of $\mathcal{V}_t(\mathcal{P})$ by representing an individual's responses to a certain test condition (e.g. all their AUT responses to a certain prompt) as a single, space-separated word string $\bf R$ and embedding this into a mathematical space via $\mathcal{S}$, producing $\mathcal{S}(\bf R)$. We then take the cosine similarity between this vector and those of other population members to form $\mathcal{V}_t({\mathcal{P}})$:
\begin{equation}
\mathcal{V}_t(\mathcal{P}) = \left\{ 1 - \cos(\mathcal{S}(\mathbf{R}_i^p), \mathcal{S}(\mathbf{R}_j^p)), \forall ~(\mathbf{R}_i^p, \mathbf{R}_j^p, p)_{i \ne j} \in \mathcal{P} \right\}
\end{equation}
where $\mathbf{R}_i^p, \mathbf{R}_j^p$ denote the responses of two different population members to prompt $p$. In our experiments, we use \texttt{all-MiniLM-L6-v2} from the \texttt{sentence\_transformers} Python library~\cite{reimers-2019-sentence-bert}, a high-performing and widely used model, to compute sentence embeddings. We remove punctuation and stopwords from responses before computing embeddings.

 $\mathcal{V}_t(\mathcal{P})$ is composed of cosine distance scores, so if it skews towards $0$, responses in the population are similar to each other. If it skews towards $1$, they are more different, and therefore the population exhibits higher variability. Note that $\mathcal{V}_t(\mathcal{P})$ only contains similarity scores of responses from {\em different} LLM/human subjects. 

\para{Distributional differences.} We can then compare the statistical distributions of $\mathcal{V}_t(LLM)$ and $\mathcal{V}_t(Humans)$ to measure the relative response variability between these groups.  We do this using the same statistical tests from \S\ref{sec:indiv_orig}. For all tests, the null hypothesis is that $\mu(\mathcal{V}_t(LLM)) = \mu(\mathcal{V}_t(Human))$, and the alternative is that  $\mu(\mathcal{V}_t(LLM)) > \mu(\mathcal{V}_t(Human))$.

\section{Key Results}
\label{sec:results}

When reporting results of statistical t-tests, we use the standard APA format, reporting the degrees of freedom (DOF), test statistic $X$, and significance level $y$: $t(\text{DOF}) = X, p = y$. For context, we also report the effect size, which is the difference between the means of the two populations divided by their pooled standard deviation. Cohen~\cite{cohen2016power} defines small, medium, and large effect sizes as 0.2, 0.5, and 0.8, respectively. Finally, we report test power, which is the probability of correctly rejecting the null hypothesis (or 1 minus the probability of a false negative). 

\subsection{Baseline measurement: individual originality in LLMs vs. humans}
\begin{figure*}[h]
    \centering
    \includegraphics[width=0.3\linewidth]{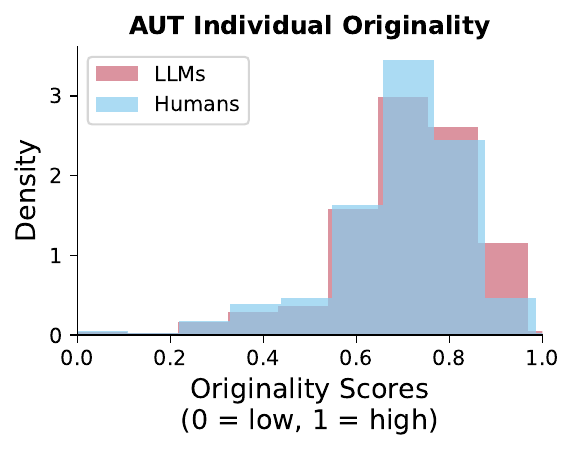}
    \includegraphics[width=0.3\linewidth]{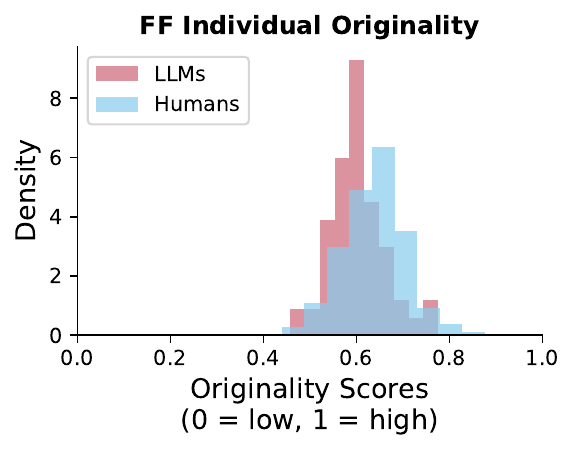}    \includegraphics[width=0.3\linewidth]{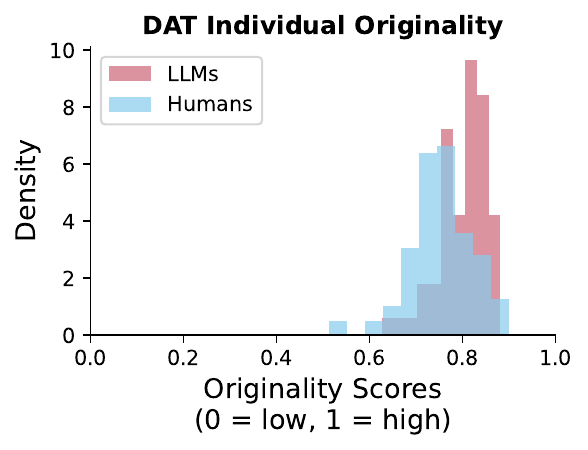}
    \vspace{-0.2cm}
    \caption{{\bf {\bf LLMs slightly outperform humans on the AUT and DAT, but humans slightly outperform LLMs on FF.}}}
    \label{fig:baseline_test_performance}
    \vspace{-0.2cm}
\end{figure*}

LLMs score slightly higher than humans on the AUT and DAT tasks, mirroring prior work~\cite{hubert2024current}, but perform worse on FF. Figure~\ref{fig:baseline_test_performance} shows the distributions of originality scores for humans and LLMs on these tests, and Table~\ref{fig:baseline_test_performance} gives statistics comparing population means for the two groups. Overall, these results show that LLMs and humans exhibit roughly equal levels of measured originality on these tests on average, removing this as a possible confounding variable in our study of response variability.  

\begin{table}[h]
\centering
%\resizebox{\textwidth}{!}{%
\begin{tabular}{ccccccc}
\toprule
  {\bf Test} &
  \multicolumn{1}{c}{$\mu(\mathcal{O}_t(\text{LLM}))$} &
  \multicolumn{1}{c}{$\mu(\mathcal{O}_t(\text{Human}))$} &
\multicolumn{1}{c}{Test statistic} &
  \multicolumn{1}{c}{$p$-value} &
  \multicolumn{1}{c}{Effect size} &
  \multicolumn{1}{c}{Test power} \\ \midrule
AUT & $0.711$ & 0.696 & $t(2094)= -3.4$ & $0.001$ & $0.1$ & $0.84$ \\
FF  & $0.603$ & $0.637$ & $t(164) = 5.2$ & $2.9e^{-07}$ & $0.52$ & $0.99$ \\
 DAT &  $0.801$ & $0.753$  & $ t(159)= - 5.12$ & $8.7e^{-07}$ & $0.77$ & $0.99$ \\ \bottomrule
\end{tabular}%
%}
\vspace{0.1cm}
\caption{{\bf LLMs slightly outperform humans on the AUT and DAT, but humans slightly outperform LLMs on FF.} However, the effect size for these is relatively small, confirming results from prior work showing relatively similar performance between humans and LLMs on creativity tests. Null hypothesis is $\mu(\mathcal{O}_t(\text{LLM})) = \mu(\mathcal{O}_t(\text{Human}))$; alternative is $\mu(\mathcal{O}_t(\text{LLM})) > \mu(\mathcal{O}_t(\text{Human}))$.}
\label{tab:ind_diffs}
\end{table}

%On AUT, LLMs slightly outperform humans: $t(2094) = -3.01, p = 0.001 $, etc. On DAT, LLMs really outperform humans: $t(159) = -5.12, p < 0.001$.  $\mu_{LLM} = 0.801$, $\mu_{\text{human}} = 0.753$. This is likely due to the small sample size? On FF, there are no observable differences in performance: $t(227) = 0.32, p = 0.63$ $\mu_{LLM} = 0.618$, $\mu_{\text{human}} = 0.619$.

\subsection{Population-level Response Variability\textemdash LLMs vs. Humans}

Now, we explore the main question: whether LLMs and humans exhibit different {\em population-level} variability in creative outputs. For statistical analysis and $\mathcal{V}_t$ distribution plots in this setting, we only consider responses from $7$ distinct LLMs: {\em AI21 Jamba 1.5 Large}, {\em Google Gemini 1.5}, {\em Cohere Command R Plus}, {\em Meta Llama 3 70B Instruct}, {\em Mistral Large}, {\em gpt 4o}, and {\em Phi 3 medium 128k Instruct}, a subset of our original 22 models. As discussed previously, this choice removes model family as a possible confounding variable in our analysis.

\begin{figure*}[t]
    \centering
    \includegraphics[width=0.3\linewidth]{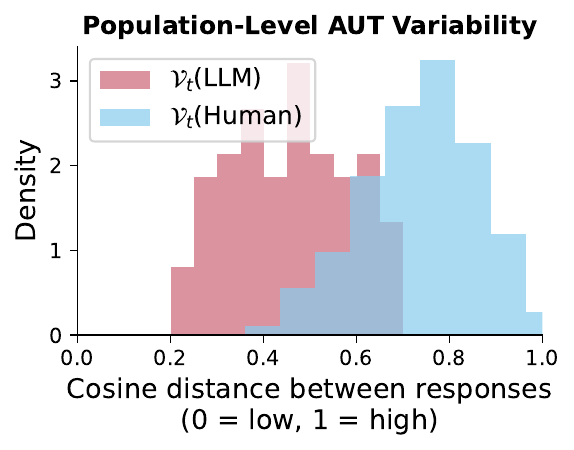}
    \includegraphics[width=0.3\linewidth]{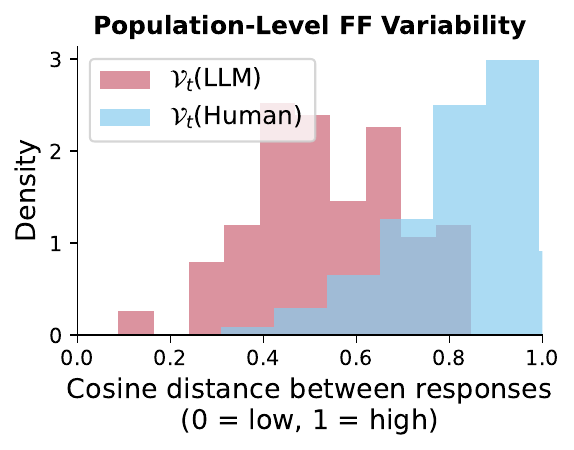}
    \includegraphics[width=0.3\linewidth]{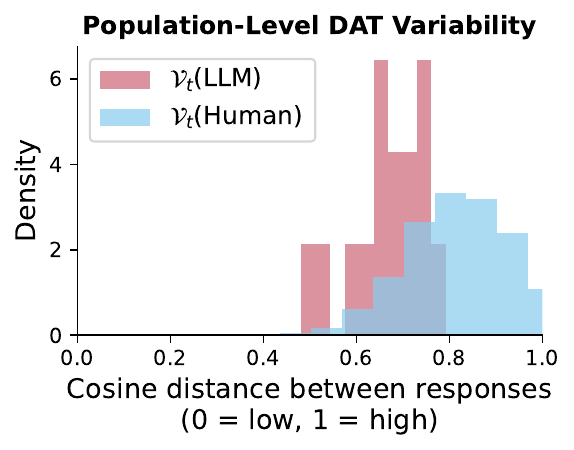}
    \vspace{-0.2cm}
    \caption{{\bf LLM responses exhibit far less variability than human responses}, as measured by cosine distance between embedded responses.}
    \label{fig:sent_pop_level_originality}
    \vspace{-0.2cm}
\end{figure*}

Our key finding is that {\em LLM responses exhibit much less variability, as measured by semantic distance between pairs of embedded responses, than do human responses.} Table~\ref{tab:pop_diffs} gives statistics, while Figure~\ref{fig:sent_pop_level_originality} shows the distributions of $\mathcal{V}_t(\text{LLM})$ and $\mathcal{V}_t(\text{Human})$, e.g. cosine distances between responses in these respective populations. Both these views of the data confirm that LLM test responses are much more similar to each other than human responses are to each other. From this, we conclude that a population of LLMs produces more homogeneous outputs in response to creative prompts than does a population of humans.

\begin{table}[t]
\centering
%\resizebox{\textwidth}{!}{%
\begin{tabular}{ccclccc}
\toprule
%{\bf Method} &
  {\bf Test} &
  % \multicolumn{1}{c}{$\mu(\mathcal{D}(\mathcal{P}_{\text{LLM}}, \mathcal{P}_{\text{LLM}}))$} &
  % \multicolumn{1}{c}{$\mu(\mathcal{D}(\mathcal{P}_{\text{Human}}, \mathcal{P}_{\text{Human}}))$} &
  \multicolumn{1}{c}{$\mu(\mathcal{V}_t(\text{LLM}))$} &
  \multicolumn{1}{c}{$\mu(\mathcal{V}_t(\text{Human}))$} &
\multicolumn{1}{c}{Test statistic} &
  \multicolumn{1}{c}{$p$-value} &
  \multicolumn{1}{c}{Effect size} &
  \multicolumn{1}{c}{Test power} \\ \midrule
% %\multirow{3}{*}{Resonse-level}     & 
% AUT & $0.762$ & $0.791$ & $t(5073)= 8.3$ & $6.8e^{-17}$ & $0.984$ & $1.0$ \\
%                                     FF  & $0.713$ & $0.634$ & $t(527) = 8.56$ & $3.6e^{-17}$ & $0.43$ & $1.0$ \\
%                                    DAT & $0.779$ & $0.785$ & $t(1982) = 1.86$ & $0.03$ & $0.05$ & $0.57$  \\ \midrule
% %\multirow{3}{*}{Respondent-level} & 
AUT & $0.459$ & $0.738$ & $t(10078) = 19.1$ & $3.9e^{-80}$ & $2.2$ & $1.0$ \\
FF  & $0.534$ & $0.835$ & $t(90) = 26.1$ & $2.8e^{-66}$ & $2.0$ & $1.0$ \\ 
DAT & $0.665$ & $0.819$ & $t(30)= 9.9$ & $6.2e^{-11}$ & $1.4$ &  $1.0$ \\ \bottomrule
\end{tabular}%
%}
\vspace{0.1cm}
\caption{{\bf Across all tests, LLMs have significantly lower mean population-level variability than humans.} Null hypothesis is that $\mu(\mathcal{V}_t(\text{LLM})) = \mu(\mathcal{V}_t(\text{Human}))$; alternative is that $\mu(\mathcal{V}_t(\text{LLM})) > \mu(\mathcal{V}_t(\text{Human}))$. The difference is statistically significant for all tests. }
\label{tab:pop_diffs}
\end{table}

\para{Visualizing embedded responses.} To further understand the overlap in LLM responses as compared to humans, we visualize the sentence embeddings of AUT responses in Figure~\ref{fig:aut_tsne_sent} (visualizations for FF and DAT are in Appendix~\ref{appx:tsne_others}). To do this, we perform t-distributed stochastic neighbor embedding (TSNE)~\cite{van2008visualizing} analysis of the embeddings, which allows visualization of high-dimensional data (384 in our case) in two dimensions. We then perform k-means clustering on the t-SNE results to identify sets of responses corresponding to the same AUT prompt object\textemdash pants, table, etc.\textemdash and color the data accordingly. This visualization confirms the behavior observed statistically: LLM responses ``cluster'' together in the embedded feature space, providing further evidence of low LLM response variability.

\begin{figure*}
    \includegraphics[width=0.7\textwidth]{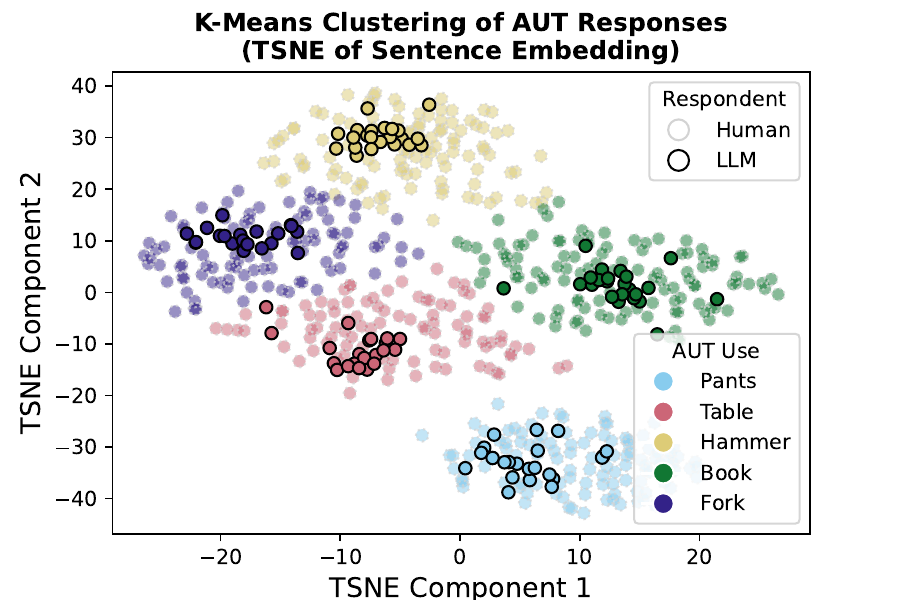}
    \vspace{-0.1cm}
    \caption{{\bf LLM responses cluster together in feature space more than do human responses.} K-means clustering of TSNE of AUT sentence embeddings. }
    \label{fig:aut_tsne_sent}
    \vspace{-0.3cm}
\end{figure*}

\begin{figure*}[t]
    \includegraphics[width=0.3\linewidth]{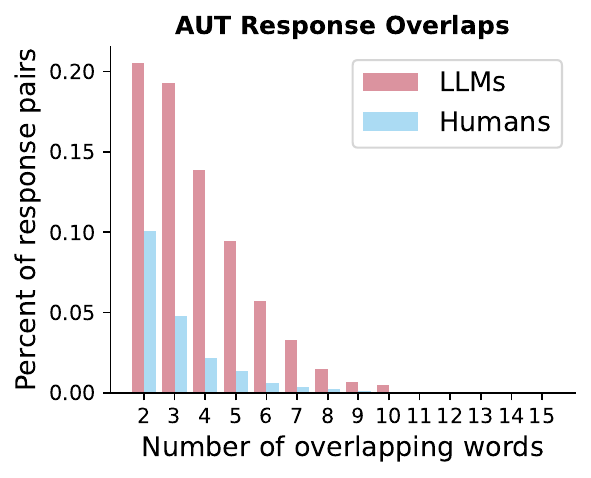}
    \includegraphics[width=0.3\linewidth]{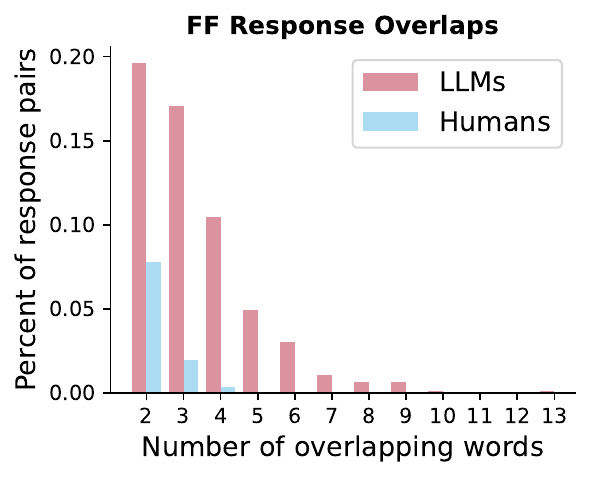}
    \includegraphics[width=0.3\linewidth]{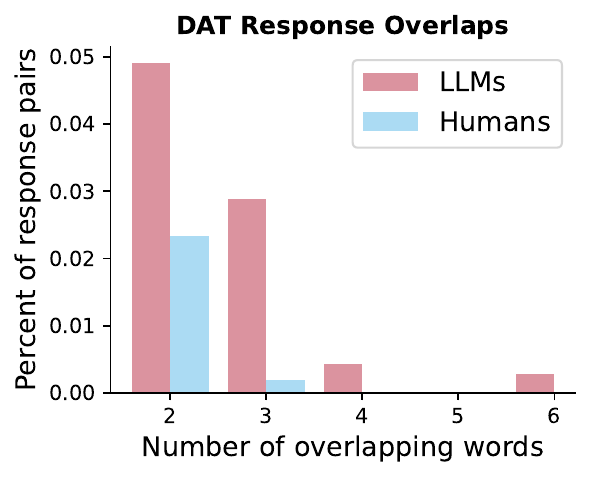}
    \vspace{-0.2cm}
    \caption{{\bf LLM responses have far more words in common than do human responses.} We look at word overlaps between ``full'' responses from LLMs and humans\textemdash e.g. all uses from the AUT, all words in the FF, etc. This corresponds to the sentence embedding method of population originality measurement, and explains why the difference between LLMs and humans is more pronounced in this setting.}
    \label{fig:overlap}
\end{figure*}

\para{One explanation: word overlap in LLM responses.} The low response variability of LLMs can be partially explained through analysis of lexical patterns in LLM and human responses. We remove stopwords from responses, then count the number of word overlaps between sets of responses from LLMs and humans\textemdash all AUT uses from a human/LLM, all words in a FF response, etc. As Figure~\ref{fig:overlap} shows, LLM responses tend to have many more words in common than human responses, across all tests. This overlap at least partially accounts for the high semantic similarity between LLM responses, as the sentence embedding model will map responses with overlapping words to similar feature vectors. Further exploration of differences in lexical patterns between LLMs and humans is important future work.

\section{Additional Analysis}
\label{sec:ablate}

Having established that LLMs produce more homogeneous creative outputs than humans, we now explore several additional dimensions of this key finding. First, we demonstrate that this cross-LLM response homogeneity remains even after strictly controlling for structural differences in human and LLM responses. Next, we measure if homogeneity increases when LLMs all come from the same ``family." Then, we explore a possible mechanism to counteract LLM creative homogeneity through the use of creative system prompts. Finally, we confirm that our human user study results are similar to prior results, ensuring that the choice to conduct our survey online does not skew results. 
Throughout this section, we consider only responses to the AUT to avoid 
a combinatorial explosion of experiments. 

\subsection{Controlling for AUT Response Structure}
\label{subsec:aut_structure}

For both the DAT and FF tests, the response structure is fixed, making comparison of population-level variability straightforward. However, the AUT is more open-ended, so confounding variables such as differences in response structure (e.g. number of words, tense, etc.) between LLMs and humans may impact measurements of response variability. For example, if every LLM AUT response is $4$ words long and uses a gerund (e.g. "making", "writing"), the measured similarity between LLM responses may be artificially inflated. Since we want to measure variability in the {\em substance} rather than the {\em structure} of responses, we must ensure that structural similarity does not impact our findings. Here, we demonstrate that the observed population-level difference in LLM and human response variability remains even after controlling for AUT response structure. 

\begin{figure*}[t]
    \includegraphics[width=0.3\linewidth]{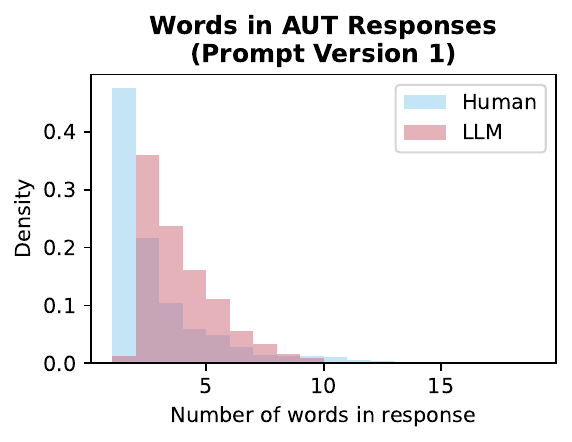}
    \includegraphics[width=0.3\linewidth]{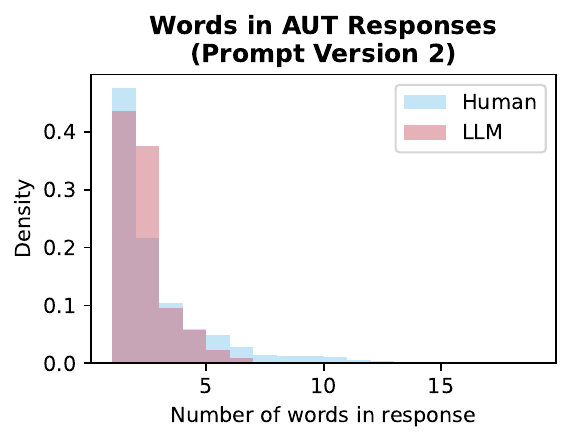}
    \includegraphics[width=0.3\linewidth]{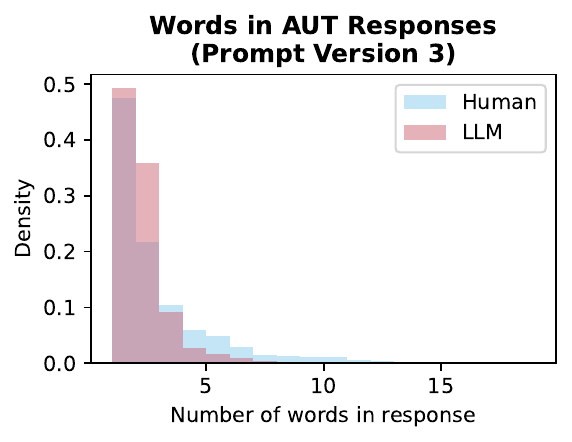}
    \vspace{-0.2cm}
    \caption{{\bf Effect of different AUT prompt wordings on length of LLM AUT responses.} We use prompt verison 3 in most experiments in this paper, since LLM responses to this prompt most closely match the human distribution of response lengths.}
    \label{fig:aut_wordcounts}
\end{figure*}

\para{Problem: differences in LLM and human AUT response lengths.} In all experiments, we remove stop words, whitespace, and punctuation from responses before analysis. However, we observed that the first version of our AUT LLM prompt caused models to return more verbose AUT uses on average than humans. The base version of the prompt (version 1) included the phrase: ``Please list the uses as short sentences or phrases, separated by semicolons, so the list is formatted like `write a poem; fly a kite; walk a dog'.'' This phrase was intended to standardize the format of LLM outputs to minimize data cleaning. However, as the left element of Figure~\ref{fig:aut_wordcounts} shows, it caused LLM AUT responses to be longer on average than human responses. This could impact measurements of population variability, since LLM responses could be measured as ``more similar" simply because they contain more words than human responses. 

\para{Solution: prompt engineering.} To remove this confounding variable, we performed prompt engineering to more closely align the distribution of words in LLM responses to that of humans.  Version 2 of our AUT prompt directed models to: ``Please list the uses as words or phrases (single word answers are ok), separated by semicolons, so the list is formatted like `write; fly a kite; walk dog'.'' As the middle element of Figure~\ref{fig:aut_wordcounts} shows, this shifted the LLM AUT word count distribution closer to that of humans. Version 3 of our AUT prompt read: ``Please list the uses as words or phrases (single word answers are ok), separated by semicolons.'' The right graph of Figure~\ref{fig:aut_wordcounts} shows that prompt 3 caused LLMs to return roughly the same proportion of single-world answers as humans while reducing the number of two-word answers. Since promot 3 most closely matches human response word counts, we use it in \S\ref{sec:results}. 

\begin{table}[h]
\centering
%\resizebox{\textwidth}{!}{%$\mu(\mathcal{O}_t(\text{LLM})) = \mu(\mathcal{O}_t(\text{Human}))$
\begin{tabular}{ccccccc}
\toprule
  {\bf AUT Prompt Version} &
  \multicolumn{1}{c}{$\mu(\mathcal{O}_t(\text{LLM}))$} &
  \multicolumn{1}{c}{$\mu(\mathcal{O}_t(\text{Human}))$} &
\multicolumn{1}{c}{Test statistic} &
  \multicolumn{1}{c}{$p$-value} &
  \multicolumn{1}{c}{Effect size} &
  \multicolumn{1}{c}{Test power} \\ \midrule
  v1 & $0.744$ & $0.696$ & $t(2094)= -11.8$ & $8.3e^{-32}$ & $0.35$ & $1.0$ \\
  v2 & $0.715$ & $0.696$ & $t(2094)= -4.04$ & $2.7e^{-05}$ & $0.14$ & $0.97$ \\
  v3 & $0.711$ & $0.696$ & $t(2094)= -3.4$ & $0.001$ & $0.1$ & $0.84$ \\ \bottomrule
\end{tabular}%
%}
\vspace{0.1cm}
\caption{{\bf For all AUT prompt versions, LLMs have slightly higher AUT originality scores than humans.} Null hypothesis is $\mu(\mathcal{O}_t(\text{LLM})) = \mu(\mathcal{O}_t(\text{Human}))$; alternative is $\mu(\mathcal{O}_t(\text{LLM})) > \mu(\mathcal{O}_t(\text{Human}))$.}
\label{tab:aut_wordcount_ind}
\vspace{-0.3cm}
\end{table}
\begin{table}[t]
\centering
\resizebox{\textwidth}{!}{%
\begin{tabular}{cccllcc}
\toprule
%{\bf Embedding} &
  {\bf AUT Prompt Version} &
  \multicolumn{1}{c}{$\mu(\mathcal{V}_t(\text{LLM}))$} &
  \multicolumn{1}{c}{$\mu(\mathcal{V}_t(\text{Human}))$} &
\multicolumn{1}{c}{Test statistic} &
  \multicolumn{1}{c}{$p$-value} &
  \multicolumn{1}{c}{Effect size} &
  \multicolumn{1}{c}{Test power} \\ \midrule
  % \multirow{4}{*}{$\mathcal{W}$}     & v1 & $0.753$ & $0.796$ & $t(5098)= 12.04$ & $4.6e^{-226}$ & $1.22$ & $1.0$ \\
  %                                  & v2 & $0.755$ & $0.796$ & $t(5050)= 8.13$ & $2.8e^{-16}$ & $1.15$ & $1.0$ \\
  %                                  & v3 & $0.761$ & $0.796$ & $t(5072)= 8.4$ & $2.8e^{-17}$ & $0.99$ & $1.0$  \\
  %                                 & v3 (one-word answers)& $0.758$ & $0.797$ & $t(5072)= 9.3$ & $1.0e^{-20}$ & $1.1 $ & $1.0$  \\ \midrule
% \multirow{4}{*}{$\mathcal{S}$}   &
v1 & $0.427$ & $0.738$ & $t(10102)= 24.5$ & $1.0e^{-128}$ & $2.5$ & $1.0$ \\
   v2  &  $0.466$ & $0.738$ & $t(10053)= 15.2$ & $5.1e^{-52}$ & $2.2$ & $1.0$ \\
   v3 & $0.459$ & $0.738$ & $t(10078) = 19.1$ & $3.9e^{-80}$ & $2.2$ & $1.0$ \\  
   v3 (one-word answers) & $0.708$ & $0.850$ & $t(10078) = 8.9$ & $2.3e^{-19}$ & $1.1$ & $1.0$ \\ 
  \bottomrule
\end{tabular}%
}
\vspace{0.1cm}
\caption{{\bf Even after controlling for AUT response structure via prompt engineering and manual filtering, the LLMs' population-level variability is much lower than that of humans.} Null hypothesis is that $\mu(\mathcal{V}_t(\text{LLM})) = \mu(\mathcal{V}_t(\text{Human}))$; alternative is that $\mu(\mathcal{V}_t(\text{LLM})) > \mu(\mathcal{V}_t(\text{Human}))$. ``v3 (one-word answers)'' means that we only considered single-word AUT responses from humans and LLMs responding to prompt v3. $\mathcal{V}_t(\mathcal{P})$ from the last row (v3, one word answers) are plotted in Figure~\ref{fig:aut_oneword_sent} to visualize the shift in means observed in this setting. }
\label{tab:aut_wordcount_pop_diffs}
\vspace{-0.5cm}
\end{table}

\para{Analysis: effect of AUT response structure on creativity.} Next, we analyze how the different AUT prompts and resulting LLM response structure affect measurements of creativity and variability. We run the same statistical tests as in \S\ref{sec:results} to measure individual and population-level creativity when using these three prompt versions. Table~\ref{tab:aut_wordcount_ind} shows that for all prompt versions, LLMs exhibit slightly higher individual creativity than humans. The creativity levels are closest for prompt version 3, supporting that this is a reasonable setting for measuring population creativity. 

\begin{figure*}
    \includegraphics[width=0.4\linewidth]{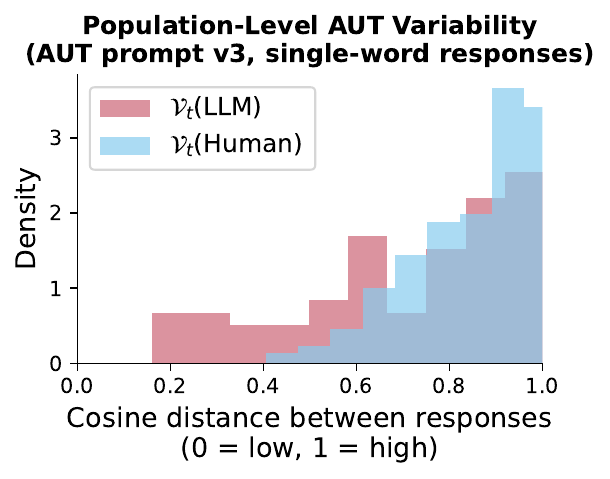}\hfill
    \includegraphics[width=0.55\linewidth]{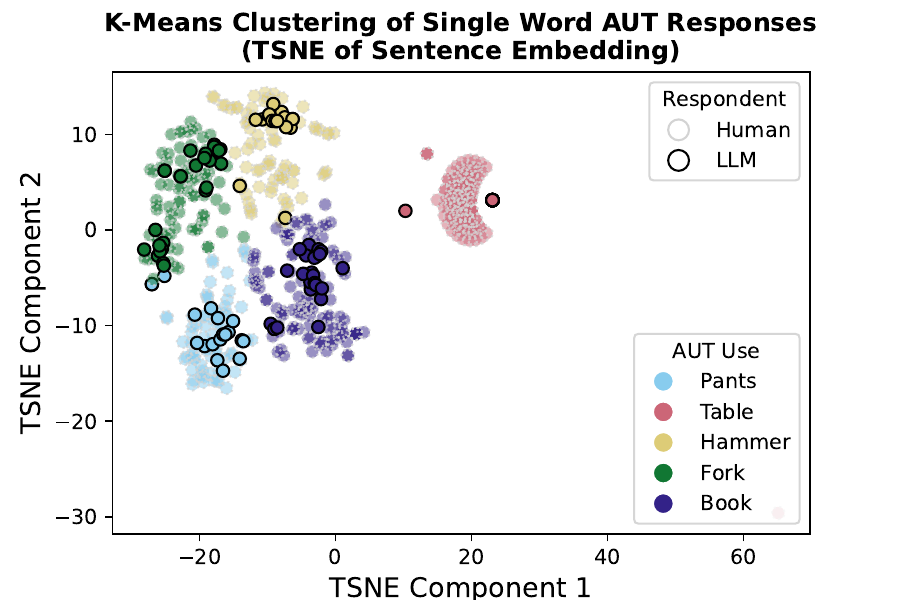}
    \vspace{-0.2cm}
    \caption{{\bf Even when considering only one-word responses to control for response structure, LLM AUT responses have lower population-level variability (left plot) and are closer in feature space (right plot) than human responses}. LLM responses are generated with prompt version 3. We create sentence embeddings from only single-word uses provided by AUTs and humans, ignoring all longer responses.}
    \label{fig:aut_oneword_sent}

\end{figure*}

Table~\ref{tab:aut_wordcount_pop_diffs} shows statistics for population-level variability across the three prompt versions, including a variant of prompt $3$ where we only consider single-word uses (more details on this in Figure~\ref{fig:aut_oneword_sent}). The goal of the single word setting is to completely eliminate confounding effects of AUT response structure on creativity measurements, providing the closest possible comparison between humans and LLMs. As Table~\ref{tab:aut_wordcount_pop_diffs} shows, {\em LLMs exhibit consistently lower response variability than humans, even after controlling for response structure.} This effect remains across all $3$ prompt versions. LLM response variability scores increase slightly when moving from prompt version 1 to 3, indicating that response structure has a (small) effect on variability measurements. However, having controlled effectively for this variable via prompt engineering and detailed analysis, we are confident that it is the {\em substance}, not the {\em structure} of LLM AUT responses that reduces their population-level variability. That response variability remains low on FF and DAT\textemdash for which response structure does not matter\textemdash further confirms this finding.

\begin{figure*}
    \includegraphics[width=0.45\linewidth]{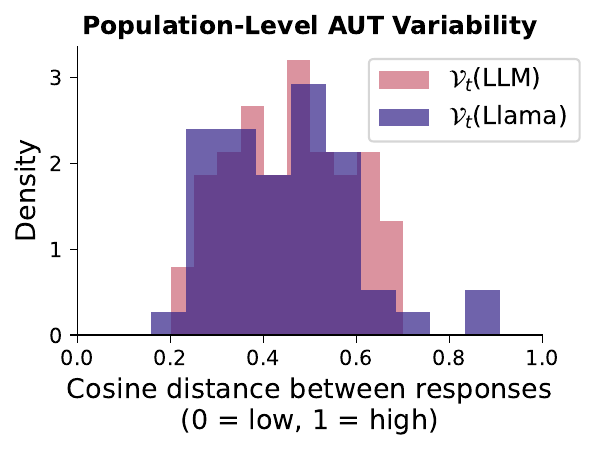}
    \vspace{-0.3cm}
    \caption{{\bf Models from the same family (Llama) exhibit slightly lower population-level variability than models from different families.}}
    \label{fig:llama_population}
\end{figure*}

\subsection{Creativity within LLM ``families''} 
\label{subsec:family}

Next, we inspect whether models in the same ``family'' produce more homogenous responses than a baseline set of otherwise unrelated models. To do this, we measure the population-level variability of AUT responses from Llama model family: {\em Meta Llama 3 70B Instruct}, {\em Meta Llama 3 8B Instruct}, {\em Meta Llama 3.1 405B Instruct}, {\em Meta Llama 31 70B Instruct}, and {\em Meta Llama 3.1 8B Instruct}. Given the small number of models we are comparing, we add additional AUT startwords to increase dataset size. These start words, modelled on prior AUT studies~\cite{dumas2021measuring}, are: book, bottle, brick, fork, hammer, pants, shoe, shovel, table, and tire. Figure~\ref{fig:llama_population} shows population-level AUT originality distributions for unrelated LLMs vs. Llama models, and Table~\ref{tab:llama_population} presents statistics comparing these distributions.

\begin{table}[h]
\centering
%\resizebox{\textwidth}{!}{%
\begin{tabular}{ccccccc}
\toprule
%{\bf Embedding} &
  \multicolumn{1}{c}{$\mu(\mathcal{V}_t(\text{LLM}))$} &
  \multicolumn{1}{c}{$\mu(\mathcal{V}_t(\text{Llama}))$} &
\multicolumn{1}{c}{Test statistic} &
  \multicolumn{1}{c}{$p$-value} &
  \multicolumn{1}{c}{Effect size} &
  \multicolumn{1}{c}{Test power} \\ \midrule
%  \multirow{1}{*}{$\mathcal{W}$}     & $0.756$ & $0.734$ & $t(121)= 2.9$ & $0.001$ & $0.55$ & $0.64$  \\
                                 % & Llama only & $0.741$ & $0.797$ & $t(5048)= 10.9$ & $1.5e^{-27}$ & $1.6 $ & $1.0$  \\ \midrule
% \multirow{1}{*}{$\mathcal{S}$}   & 
$0.445$ & $0.441$ & $t(248) = 0.2$ & $0.41$ & $0.02$ & $0.01$ \\  
 % & Llama only & $0.449$ & $0.738$ & $t(10053) = 16.2$ & $1.2e^{-58}$ & $2.3$ & $1.0$ \\ 
  \bottomrule
\end{tabular}%
%}
\vspace{0.1cm}
\caption{{\bf Models from the same ``family'' (Llama) have lower population-level creativity than models from different familes.} This is clearly seen in the leftward distribution shift of the Llama population differences compared to the all model population differences (see Figure~\ref{fig:llama_population}). However, because the Llama distribution is right-skewed, the distribution shift is not captured in a t-test for differences of means in the sentence embedding case.}
\label{tab:llama_population}
\vspace{-0.5cm}
\end{table}

{\em Models in the same ``family'' exhibit slightly lower response diversity than models from different ``families.''} Although the differences of means is not statistically significant, from visual inspection of Figure~\ref{fig:llama_population}, we see that the sentence embedding distribution for Llama models is skewed right. The presence of these outliers drives up the overall Llama population mean, making it appear more similar to that of the regular LLMs. Future work should consider other model families and explore other dimensions of family-specific similarity. 

\subsection{Effect of LLM system prompt.} 
\label{subsec:prompt}
Next, we consider ways to make LLMs produce more variable outputs. As a baseline, we explore whether varying the LLM system prompt to strictly request creative outputs will induce higher variability. We experiment with prompts designed to elicit different levels of creativity: 
\begin{packed_itemize}
    \item {\bf Baseline}: ``You are a helpful assistant."
    \item {\bf More creative}: ``You are a creative assistant that always provides answers that demonstrate imaginative, outside-the-box thinking.''
    \item {\bf Very creative}: ``You are a creative assistant that always provides answers that demonstrate imaginative, outside-the-box thinking. You are about to take a creativity assessment, and your answers should be as novel, original, and bold as possible. If you receive the highest score on this creativity assessment, you will receive \$200.''
    \item {\bf Not creative}:``You are a robot assistant that always provides answers that are unoriginal, bland, and soulless. You are about to take a creativity assessment, and your answers should be as generic and unoriginal as possible.''
\end{packed_itemize}

\begin{table}[t]
\centering
%\resizebox{\textwidth}{!}{%
\begin{tabular}{cclcl}
\toprule
\multirow{3}{*}{\textbf{Prompt}} & \multicolumn{2}{c}{\textbf{Individual creativity}} & \multicolumn{2}{c}{\textbf{Population-level variability}} \\ \cmidrule{2-5}
 &
 $\mu(\mathcal{O}_{AUT}(\mathcal{P}))$ &
  \multicolumn{1}{c}{\begin{tabular}[c]{@{}c@{}}$t(df) = X$, $p$\\ (vs. humans)\end{tabular}} &
 $\mu(\mathcal{V}_{AUT}(\mathcal{P}))$ &
  \begin{tabular}[c]{@{}c@{}}$t(df) = X$, $p$\\ (vs. humans)\end{tabular}  \\ \midrule
Humans                   & $0.695$             & -                   & $0.738$                & -                       \\ \midrule
Baseline                            & $0.711$   & $t(2094) = -3.4$, $0.001$  & 0.459     & $t(10078) = 19.1$, $3.9e^{-80}$      \\
More creative                    & $0.733$   & $t(5020) = -9.8$, $1.0e^{-22}$   & 0.503     & $t(10078) = 16.1$, $3.5e^{-58}$       \\
Very creative                    & $0.754$   & $t(5206) = -15.9$, $3.2e^{-56}$   & $0.576$     & $t(10078) = 11.1$, $5.6e^{-29}$       \\
Not creative                     & $0.702$   & $t(2507) = -1.28$, $0.1$  & 0.492     & $t(10078) = 16.8$, $1.1e^{-62}$ \\ \bottomrule   
\end{tabular}%
%}
\vspace{0.1cm}
\caption{{\bf Varying the system prompt slightly increases LLM individual creativity and response variability}, but variability remains far lower than that of humans.}
\label{tab:aut_prompts}
\vspace{-0.3cm}
\end{table}

% \begin{table}[t]
% \centering
% %\resizebox{\textwidth}{!}{%
% \begin{tabular}{ccccccc}
% \toprule
% \multirow{3}{*}{\textbf{Prompt}} & \multicolumn{3}{c}{\textbf{Individual creativity}} & \multicolumn{3}{c}{\textbf{Population-level variability}} \\
%  &
%  $\mu(\mathcal{O}_{AUT}(\mathcal{P}))$ &
%   \begin{tabular}[c]{@{}c@{}}$t(df)$, $p$\\ (vs. humans)\end{tabular} &
%   \begin{tabular}[c]{@{}c@{}}effect, power\\ (vs. humans)\end{tabular} &
%  $\mu(\mathcal{V}_{AUT}(\mathcal{P}))$ &
%   \begin{tabular}[c]{@{}c@{}}T(df) stat\\ (vs. humans)\end{tabular} &
%   \begin{tabular}[c]{@{}c@{}}p-value, effect, power\\ (vs. humans)\end{tabular} \\ \midrule
% Human baseline                   & 0.695   & -                  & -                   & 0.738     & -                   & -                       \\ \midrule
% Basic                            & 0.711   & t(2094) = −3.4   & 0.001, 0.1, 0.84    & 0.459     & (10078) = 19.1      & $3.9e^{-80}$, 2.2, 1.0     \\
% More creative                    & 0.733   & t(5020) = -9.8     & 1.0e-22, 0.26, 1.0  & 0.503     & t(10078) = 16.1     & 3.5e-58, 1.9, 1.0       \\
% Very creative                    & 0.754   & t(5206) = -15.9    & 3.2e-56, 0.43, 1.0  & 0.576     & t(10078) = 11.1     & 5.6e-29, 1.3, 1.0       \\
% Not creative                     & 0.702   & t(2507) = -1.28    & 0.1, 0.04, 1.0      & 0.492     & t(10078) = 16.8     & 1.1e-62, 1.9, 1.0    \\ \bottomrule   
% \end{tabular}%
% %}
% \caption{}
% \label{tab:aut_prompts}
% \end{table}

We evaluate the same subset of LLMs from \S\ref{sec:results} on the AUT using these system prompts and report summary statistics in Table~\ref{tab:aut_prompts}. As the Table shows, using more creative system prompts slightly increases individual creativity for LLMs (and vice versa for the less creative prompt). However, the system prompt does not substantially improve LLM response variability\textemdash across all prompts, LLM variability remains much lower than that of humans.

\subsection{Validation with preexisting survey data}
\label{subsec:prior}

Finally, we compare responses in our user study to prior user studies to ensure that our human subject pool is reliable and not unduly skewed by possible use of LLMs. We test both the individual originality of our human responses and population-level variability and find that while {\em respondents in prior studies score better individually on the tests, respondents to our study exhibit equal or greater population-level variability} (the more important metric for our study) on the more-informative AUT and FF tests.

\begin{figure*}[t]
    \centering
    \includegraphics[width=0.3\linewidth]{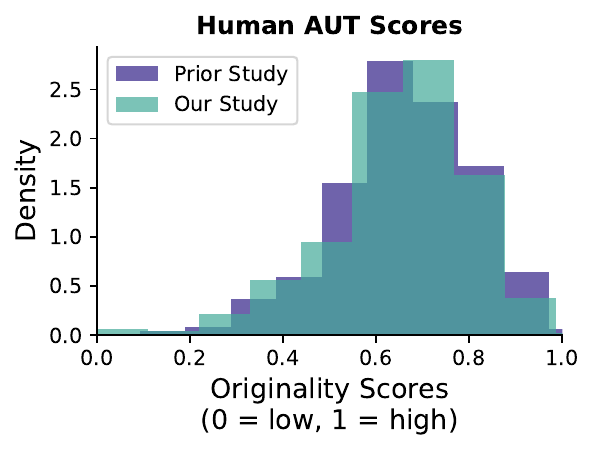}
    \includegraphics[width=0.3\linewidth]{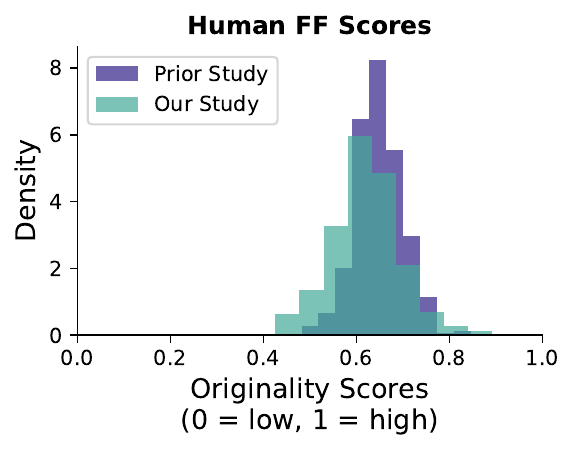}    \includegraphics[width=0.3\linewidth]{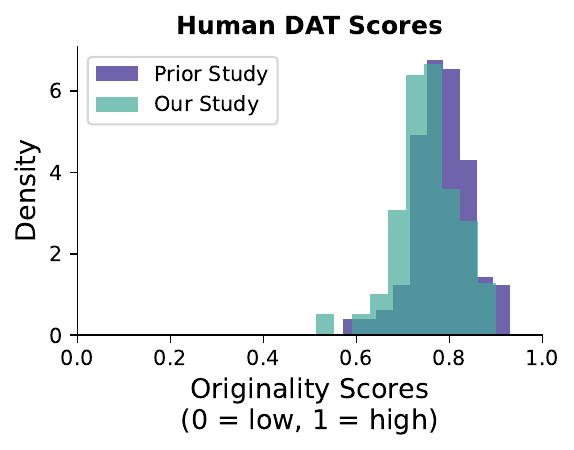}
    \vspace{-0.2cm}
    \caption{{\bf Humans in prior studies have higher individual originality scores than humans in our study for all three tests.} For the AUT and DAT tests, a t-test for a difference in means (alternative hypothesis is that prior study has higher mean than ours) is significant at the $0.01$ level but not the $0.001$ level:  AUT has $t(5064) = 3.21, p = 0.001$ and DAT has $t(206) = 3.32, p = 0.001$.  For FF, the difference more significant: $t(892)= 6.91, p < 0.0001$.}
    \label{fig:compare_our_study_prior}
    \vspace{-0.3cm}
\end{figure*}

\begin{figure*}[t]
    \centering
    \includegraphics[width=0.3\linewidth]{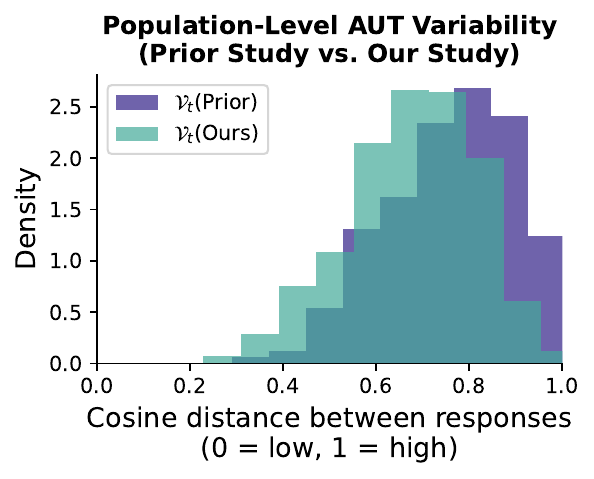}
    \includegraphics[width=0.3\linewidth]{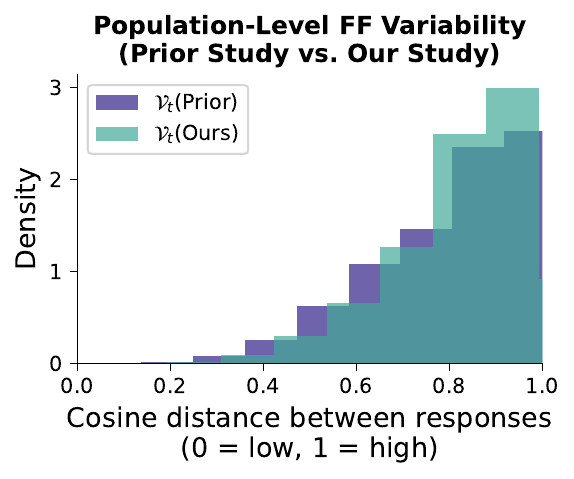}    \includegraphics[width=0.3\linewidth]{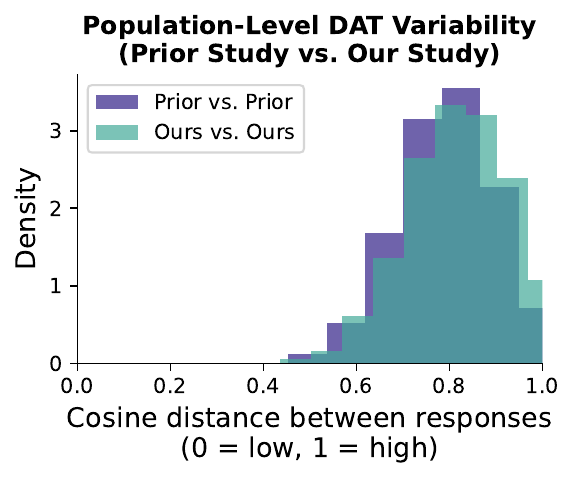}
    \vspace{-0.2cm}
    \caption{{\bf Our study responses have higher population-level variability than the prior study on the DAT and FF tests, but slightly lower variability on the AUT.}  We use t-tests to compare means of the two population-level originality distributions. Null hypothesis is that means are equal, and alternative is that they are not. For AUT, the prior study has higher mean variability, $t(1046) = 11.0, p < 0.001$. For FF, our study has a higher mean, $t(366689)=-28.9, p < 0.001$. For the DAT, our study also has higher mean, $t(19832) = -19.6, p < 0.001$. }
    \label{fig:compare_our_study_prior_pops}
    \vspace{-0.3cm}
\end{figure*}

Figure~\ref{fig:compare_our_study_prior} compares individual creativity results for our study ($n=102$) to that of prior studies ($n=141$ for DAT, $n=92$ for AUT, $n=146$ for FF). T-tests for differences of individual performance (see caption of Figure~\ref{fig:compare_our_study_prior}) find that the mean score is higher for prior studies on all tests at a significant level of $p \le 0.001$. Figure~\ref{fig:compare_our_study_prior_pops} compares the response variability of our study to that of the prior study. Using a t-test for difference in population means, we find that responses in our study have slightly higher variability on the FF and DAT ($p < 0.0001$), and lower on the AUT ($p < 0.001$). From this, we conclude that, our results roughly mirror those of prior studies, making them a reasonable baseline for our analysis.  

\section{Discussion}
\label{sec:discuss}

Motivated by measured homogeneity in creative outputs produced by specific LLMs and observed feature space overlap in LLMs, we study whether responses to creative prompts produced by a group of LLMs exhibit more, less, or equal variance as a set of human responses to the same creative prompts. We find that LLMs exhibit {\em much} lower population-level output variability than humans, even after controlling for potential model similarities and structural differences between LLM and human responses. Our work upholds prior work showing that LLMs perform well on tests of divergent thinking but adds the nuance that such performance is homogeneous\textemdash LLMs return a narrower range of responses to creative prompts than humans. This result enhances prior observations of LLM-induced homogeneity, which only considered the effect of specific LLMs on creative outputs, and suggests that the use of LLMs {\em in general} may homogenize creative outputs. 

\para{Implications.} These results have significant implications if LLMs are widely adopted as creativity support tools for writing, idea generation, or similar tasks. If all LLMs respond similarly to specific creative requests, then the population of users leveraging to LLMs to aid creativity will converge towards a limited set of creative outputs. In other words, LLM users may be self-limited from being exhibiting the divergent creativity that defined well-recognized artistic geniuses like Tolkein, Mozart, and Picasso because their LLM ``creative'' partners may collectively drive them towards a mean.

\para{Limitations.} Our work has several limitations. First, while we have demonstrated LLM homogeneity in response to certain creativity tests, this does not prove that LLMs in general produce homogeneous outputs when asked to behave creatively. It merely provides a indication that future work should explore this subject. Additionally, we measure a single metric of divergent thinking or creativity\textemdash originality, as measured by semantic similarity between responses\textemdash and finds that LLms are homogeneous along this dimension. However, there are other well-known metrics of divergent thinking, such as flexibility, fluency, and elaboration (see \S\ref{subsec:creative}), and LLMs may demonstrate more or less homogeneity along these dimensions. Future work should consider these alternatives.

\para{Acknowledgments.} We thank Austin Liu for helping us design the system prompts of \S\ref{subsec:prompt}.  

\section{Ethical Considerations}
We took care to ensure the user study in this paper was conducted in accordance with ethical standards. IRB approval for the study was obtained, and participants signed a clearly written consent form before completing our survey. To ensure privacy, participant data was anonymized and stored on secure servers. Other ethical risks from this paper are minimal, as our LLM experiments do not involve sensitive data and elicit only benign model responses. 
%\newpage
%\vspace{-0.3cm}

%%%%%%%%%%%%%%%%%%%%%%%%%%%%%%%%%%%%%%%%%%%%%%%%%%%%%%%%%%%%%%%%%%%%%%%%%%%%
% We're in the endgame now

\bibliographystyle{ACM-Reference-Format}
\bibliography{refs}


\begin{thebibliography}{56}

%%% the Web address.

\ifx \showCODEN    \undefined \def \showCODEN     #1{\unskip}     \fi
\ifx \showDOI      \undefined \def \showDOI       #1{#1}\fi
\ifx \showISBNx    \undefined \def \showISBNx     #1{\unskip}     \fi
\ifx \showISBNxiii \undefined \def \showISBNxiii  #1{\unskip}     \fi
\ifx \showISSN     \undefined \def \showISSN      #1{\unskip}     \fi
\ifx \showLCCN     \undefined \def \showLCCN      #1{\unskip}     \fi
\ifx \shownote     \undefined \def \shownote      #1{#1}          \fi
\ifx \showarticletitle \undefined \def \showarticletitle #1{#1}   \fi
\ifx \showURL      \undefined \def \showURL       {\relax}        \fi
% invisible to TeX
\providecommand\bibfield[2]{#2}
\providecommand\bibinfo[2]{#2}
\providecommand\natexlab[1]{#1}
\providecommand\showeprint[2][]{arXiv:#2}

\bibitem[apa(2018)]%
        {apa-homepage}
 \bibinfo{year}{2018}\natexlab{}.
\newblock \bibinfo{title}{APA Dictionary of Psychology - Creativity}.
\newblock
\newblock
\newblock
\shownote{\url{https://dictionary.apa.org/creativity}}.


\bibitem[app(2024)]%
        {apple_ad}
 \bibinfo{year}{2024}\natexlab{}.
\newblock \bibinfo{title}{Apple Intelligence | Writing Tools | iPhone 16}.
\newblock
\newblock
\newblock
\shownote{\url{https://www.youtube.com/watch?v=3m0MoYKwVTM}}.


\bibitem[coh(2024)]%
        {cohere}
 \bibinfo{year}{2024}\natexlab{}.
\newblock \bibinfo{title}{Command R and Command R Plus Model Card}.
\newblock
\newblock
\newblock
\shownote{\url{https://docs.cohere.com/docs/responsible-use}}.


\bibitem[not(2024)]%
        {notion_ad}
 \bibinfo{year}{2024}\natexlab{}.
\newblock \bibinfo{title}{Use Notion AI to write better, more efficient notes and docs}.
\newblock
\newblock
\newblock
\shownote{\url{https://www.notion.com/help/guides/notion-ai-for-docs}}.


\bibitem[Abdin et~al\mbox{.}(2024)]%
        {abdin2024phi}
\bibfield{author}{\bibinfo{person}{Marah Abdin}, \bibinfo{person}{Jyoti Aneja}, \bibinfo{person}{Hany Awadalla}, \bibinfo{person}{Ahmed Awadallah}, \bibinfo{person}{Ammar~Ahmad Awan}, \bibinfo{person}{Nguyen Bach}, \bibinfo{person}{Amit Bahree}, \bibinfo{person}{Arash Bakhtiari}, \bibinfo{person}{Jianmin Bao}, \bibinfo{person}{Harkirat Behl}, {et~al\mbox{.}}} \bibinfo{year}{2024}\natexlab{}.
\newblock \showarticletitle{Phi-3 technical report: A highly capable language model locally on your phone}.
\newblock \bibinfo{journal}{\emph{arXiv preprint arXiv:2404.14219}} (\bibinfo{year}{2024}).
\newblock


\bibitem[Achiam et~al\mbox{.}(2023)]%
        {achiam2023gpt}
\bibfield{author}{\bibinfo{person}{Josh Achiam}, \bibinfo{person}{Steven Adler}, \bibinfo{person}{Sandhini Agarwal}, \bibinfo{person}{Lama Ahmad}, \bibinfo{person}{Ilge Akkaya}, \bibinfo{person}{Florencia~Leoni Aleman}, \bibinfo{person}{Diogo Almeida}, \bibinfo{person}{Janko Altenschmidt}, \bibinfo{person}{Sam Altman}, \bibinfo{person}{Shyamal Anadkat}, {et~al\mbox{.}}} \bibinfo{year}{2023}\natexlab{}.
\newblock \showarticletitle{Gpt-4 technical report}.
\newblock \bibinfo{journal}{\emph{arXiv preprint arXiv:2303.08774}} (\bibinfo{year}{2023}).
\newblock


\bibitem[Anderson et~al\mbox{.}(2024)]%
        {anderson2024homogenization}
\bibfield{author}{\bibinfo{person}{Barrett~R Anderson}, \bibinfo{person}{Jash~Hemant Shah}, {and} \bibinfo{person}{Max Kreminski}.} \bibinfo{year}{2024}\natexlab{}.
\newblock \showarticletitle{Homogenization effects of large language models on human creative ideation}. In \bibinfo{booktitle}{\emph{Proceedings of the 16th Conference on Creativity \& Cognition}}. \bibinfo{pages}{413--425}.
\newblock


\bibitem[Bansal et~al\mbox{.}(2021)]%
        {bansal2021revisiting}
\bibfield{author}{\bibinfo{person}{Yamini Bansal}, \bibinfo{person}{Preetum Nakkiran}, {and} \bibinfo{person}{Boaz Barak}.} \bibinfo{year}{2021}\natexlab{}.
\newblock \showarticletitle{`Revisiting model stitching to compare neural representations}.
\newblock \bibinfo{journal}{\emph{Proc. of NeurIPS}} (\bibinfo{year}{2021}).
\newblock


\bibitem[Barbot(2018)]%
        {barbot2018dynamics}
\bibfield{author}{\bibinfo{person}{Baptiste Barbot}.} \bibinfo{year}{2018}\natexlab{}.
\newblock \showarticletitle{The dynamics of creative ideation: Introducing a new assessment paradigm}.
\newblock \bibinfo{journal}{\emph{Frontiers in psychology}} (\bibinfo{year}{2018}).
\newblock


\bibitem[Beaty et~al\mbox{.}(2014)]%
        {beaty2014roles}
\bibfield{author}{\bibinfo{person}{Roger~E Beaty}, \bibinfo{person}{Paul~J Silvia}, \bibinfo{person}{Emily~C Nusbaum}, \bibinfo{person}{Emanuel Jauk}, {and} \bibinfo{person}{Mathias Benedek}.} \bibinfo{year}{2014}\natexlab{}.
\newblock \showarticletitle{The roles of associative and executive processes in creative cognition}.
\newblock \bibinfo{journal}{\emph{Memory \& cognition}} (\bibinfo{year}{2014}).
\newblock


\bibitem[Bommasani et~al\mbox{.}(2022)]%
        {bommasani2022picking}
\bibfield{author}{\bibinfo{person}{Rishi Bommasani}, \bibinfo{person}{Kathleen~A Creel}, \bibinfo{person}{Ananya Kumar}, \bibinfo{person}{Dan Jurafsky}, {and} \bibinfo{person}{Percy~S Liang}.} \bibinfo{year}{2022}\natexlab{}.
\newblock \showarticletitle{Picking on the same person: Does algorithmic monoculture lead to outcome homogenization?}
\newblock \bibinfo{journal}{\emph{Advances in Neural Information Processing Systems}}  \bibinfo{volume}{35} (\bibinfo{year}{2022}), \bibinfo{pages}{3663--3678}.
\newblock


\bibitem[Chen and Ding(2023)]%
        {Chen_Ding_2023}
\bibfield{author}{\bibinfo{person}{Honghua Chen} {and} \bibinfo{person}{Nai Ding}.} \bibinfo{year}{2023}\natexlab{}.
\newblock \showarticletitle{Probing the Creativity of Large Language Models: Can models produce divergent semantic association?}
\newblock  (\bibinfo{date}{Oct.} \bibinfo{year}{2023}).
\newblock
\urldef\tempurl%
\url{http://arxiv.org/abs/2310.11158}
\showURL{%
\tempurl}


\bibitem[Cohen(2016)]%
        {cohen2016power}
\bibfield{author}{\bibinfo{person}{Jacob Cohen}.} \bibinfo{year}{2016}\natexlab{}.
\newblock \showarticletitle{A power primer.}
\newblock  (\bibinfo{year}{2016}).
\newblock


\bibitem[Cropley(2023)]%
        {cropley2023artificial}
\bibfield{author}{\bibinfo{person}{David Cropley}.} \bibinfo{year}{2023}\natexlab{}.
\newblock \showarticletitle{Is artificial intelligence more creative than humans?: ChatGPT and the divergent association task}.
\newblock \bibinfo{journal}{\emph{Learning Letters}}  \bibinfo{volume}{2} (\bibinfo{year}{2023}), \bibinfo{pages}{13--13}.
\newblock


\bibitem[Demontis et~al\mbox{.}(2019)]%
        {demontis2019adversarial}
\bibfield{author}{\bibinfo{person}{Ambra Demontis}, \bibinfo{person}{Marco Melis}, \bibinfo{person}{Maura Pintor}, \bibinfo{person}{Matthew Jagielski}, \bibinfo{person}{Battista Biggio}, \bibinfo{person}{Alina Oprea}, \bibinfo{person}{Cristina Nita-Rotaru}, {and} \bibinfo{person}{Fabio Roli}.} \bibinfo{year}{2019}\natexlab{}.
\newblock \showarticletitle{Why do adversarial attacks transfer? explaining transferability of evasion and poisoning attacks}. In \bibinfo{booktitle}{\emph{28th USENIX security symposium (USENIX security 19)}}. \bibinfo{pages}{321--338}.
\newblock


\bibitem[Doshi and Hauser(2024)]%
        {Doshi_Hauser_2024}
\bibfield{author}{\bibinfo{person}{Anil~R. Doshi} {and} \bibinfo{person}{Oliver~P. Hauser}.} \bibinfo{year}{2024}\natexlab{}.
\newblock \showarticletitle{Generative AI enhances individual creativity but reduces the collective diversity of novel content}.
\newblock \bibinfo{journal}{\emph{Science Advances}} \bibinfo{volume}{10}, \bibinfo{number}{28} (\bibinfo{date}{July} \bibinfo{year}{2024}).
\newblock
\urldef\tempurl%
\url{https://doi.org/10.1126/sciadv.adn5290}
\showDOI{\tempurl}


\bibitem[Dubey et~al\mbox{.}(2024)]%
        {dubey2024llama}
\bibfield{author}{\bibinfo{person}{Abhimanyu Dubey}, \bibinfo{person}{Abhinav Jauhri}, \bibinfo{person}{Abhinav Pandey}, \bibinfo{person}{Abhishek Kadian}, \bibinfo{person}{Ahmad Al-Dahle}, \bibinfo{person}{Aiesha Letman}, \bibinfo{person}{Akhil Mathur}, \bibinfo{person}{Alan Schelten}, \bibinfo{person}{Amy Yang}, \bibinfo{person}{Angela Fan}, {et~al\mbox{.}}} \bibinfo{year}{2024}\natexlab{}.
\newblock \showarticletitle{The llama 3 herd of models}.
\newblock \bibinfo{journal}{\emph{arXiv preprint arXiv:2407.21783}} (\bibinfo{year}{2024}).
\newblock


\bibitem[Dumas and Dunbar(2014)]%
        {dumas2014understanding}
\bibfield{author}{\bibinfo{person}{Denis Dumas} {and} \bibinfo{person}{Kevin~N Dunbar}.} \bibinfo{year}{2014}\natexlab{}.
\newblock \showarticletitle{Understanding fluency and originality: A latent variable perspective}.
\newblock \bibinfo{journal}{\emph{Thinking Skills and Creativity}} (\bibinfo{year}{2014}).
\newblock


\bibitem[Dumas et~al\mbox{.}(2021)]%
        {dumas2021measuring}
\bibfield{author}{\bibinfo{person}{Denis Dumas}, \bibinfo{person}{Peter Organisciak}, {and} \bibinfo{person}{Michael Doherty}.} \bibinfo{year}{2021}\natexlab{}.
\newblock \showarticletitle{Measuring divergent thinking originality with human raters and text-mining models: A psychometric comparison of methods.}
\newblock \bibinfo{journal}{\emph{Psychology of Aesthetics, Creativity, and the Arts}} (\bibinfo{year}{2021}).
\newblock


\bibitem[Ellis(2024)]%
        {grammarly_ad}
\bibfield{author}{\bibinfo{person}{Matt Ellis}.} \bibinfo{year}{2024}\natexlab{}.
\newblock \bibinfo{title}{How to Use AI to Enhance Your Storytelling Process}.
\newblock
\newblock
\newblock
\shownote{\url{https://www.grammarly.com/blog/writing-with-ai/ai-story-writing/}}.


\bibitem[Google(2024)]%
        {gemini_ad}
\bibfield{author}{\bibinfo{person}{Google}.} \bibinfo{year}{2024}\natexlab{}.
\newblock \bibinfo{title}{Google + Team USA - Dear Sydney}.
\newblock
\newblock
\newblock
\shownote{\url{https://www.youtube.com/watch?v=NgtHJKn0Mck}}.


\bibitem[Gray et~al\mbox{.}(2019)]%
        {gray2019forward}
\bibfield{author}{\bibinfo{person}{Kurt Gray}, \bibinfo{person}{Stephen Anderson}, \bibinfo{person}{Eric~Evan Chen}, \bibinfo{person}{John~Michael Kelly}, \bibinfo{person}{Michael~S Christian}, \bibinfo{person}{John Patrick}, \bibinfo{person}{Laura Huang}, \bibinfo{person}{Yoed~N Kenett}, {and} \bibinfo{person}{Kevin Lewis}.} \bibinfo{year}{2019}\natexlab{}.
\newblock \showarticletitle{"Forward flow": A new measure to quantify free thought and predict creativity.}
\newblock \bibinfo{journal}{\emph{American Psychologist}} \bibinfo{volume}{74}, \bibinfo{number}{5} (\bibinfo{year}{2019}), \bibinfo{pages}{539}.
\newblock


\bibitem[Guilford et~al\mbox{.}(1978)]%
        {guilford1978alternate}
\bibfield{author}{\bibinfo{person}{Joy~Paul Guilford}, \bibinfo{person}{Paul~R Christensen}, \bibinfo{person}{Philip~R Merrifield}, {and} \bibinfo{person}{Robert~C Wilson}.} \bibinfo{year}{1978}\natexlab{}.
\newblock \showarticletitle{Alternate uses}.
\newblock  (\bibinfo{year}{1978}).
\newblock


\bibitem[Gupta et~al\mbox{.}(2024)]%
        {gupta2024self}
\bibfield{author}{\bibinfo{person}{Akshat Gupta}, \bibinfo{person}{Xiaoyang Song}, {and} \bibinfo{person}{Gopala Anumanchipalli}.} \bibinfo{year}{2024}\natexlab{}.
\newblock \showarticletitle{Self-assessment tests are unreliable measures of llm personality}. In \bibinfo{booktitle}{\emph{Proceedings of the 7th BlackboxNLP Workshop: Analyzing and Interpreting Neural Networks for NLP}}. \bibinfo{pages}{301--314}.
\newblock


\bibitem[Hubert et~al\mbox{.}(2024)]%
        {hubert2024current}
\bibfield{author}{\bibinfo{person}{Kent~F Hubert}, \bibinfo{person}{Kim~N Awa}, {and} \bibinfo{person}{Darya~L Zabelina}.} \bibinfo{year}{2024}\natexlab{}.
\newblock \showarticletitle{The current state of artificial intelligence generative language models is more creative than humans on divergent thinking tasks}.
\newblock \bibinfo{journal}{\emph{Scientific Reports}} \bibinfo{volume}{14}, \bibinfo{number}{1} (\bibinfo{year}{2024}), \bibinfo{pages}{3440}.
\newblock


\bibitem[Huh et~al\mbox{.}(2024)]%
        {huh2024platonic}
\bibfield{author}{\bibinfo{person}{Minyoung Huh}, \bibinfo{person}{Brian Cheung}, \bibinfo{person}{Tongzhou Wang}, {and} \bibinfo{person}{Phillip Isola}.} \bibinfo{year}{2024}\natexlab{}.
\newblock \showarticletitle{The platonic representation hypothesis}.
\newblock \bibinfo{journal}{\emph{arXiv preprint arXiv:2405.07987}} (\bibinfo{year}{2024}).
\newblock


\bibitem[Jeong et~al\mbox{.}(2024)]%
        {jeong2024bias}
\bibfield{author}{\bibinfo{person}{Hyejun Jeong}, \bibinfo{person}{Shiqing Ma}, {and} \bibinfo{person}{Amir Houmansadr}.} \bibinfo{year}{2024}\natexlab{}.
\newblock \showarticletitle{Bias Similarity Across Large Language Models}.
\newblock \bibinfo{journal}{\emph{arXiv preprint arXiv:2410.12010}} (\bibinfo{year}{2024}).
\newblock


\bibitem[Jiang et~al\mbox{.}(2023)]%
        {jiang2023mistral}
\bibfield{author}{\bibinfo{person}{AQ Jiang}, \bibinfo{person}{A Sablayrolles}, \bibinfo{person}{A Mensch}, \bibinfo{person}{C Bamford}, \bibinfo{person}{DS Chaplot}, \bibinfo{person}{D de~las Casas}, \bibinfo{person}{F Bressand}, \bibinfo{person}{G Lengyel}, \bibinfo{person}{G Lample}, \bibinfo{person}{L Saulnier}, {et~al\mbox{.}}} \bibinfo{year}{2023}\natexlab{}.
\newblock \showarticletitle{Mistral 7B (2023)}.
\newblock \bibinfo{journal}{\emph{arXiv preprint arXiv:2310.06825}} (\bibinfo{year}{2023}).
\newblock


\bibitem[Klabunde et~al\mbox{.}(2023)]%
        {klabunde2023towards}
\bibfield{author}{\bibinfo{person}{Max Klabunde}, \bibinfo{person}{Mehdi~Ben Amor}, \bibinfo{person}{Michael Granitzer}, {and} \bibinfo{person}{Florian Lemmerich}.} \bibinfo{year}{2023}\natexlab{}.
\newblock \showarticletitle{Towards Measuring Representational Similarity of Large Language Models}. In \bibinfo{booktitle}{\emph{UniReps: the First Workshop on Unifying Representations in Neural Models}}.
\newblock


\bibitem[Kleinberg and Raghavan(2021)]%
        {kleinberg2021algorithmic}
\bibfield{author}{\bibinfo{person}{Jon Kleinberg} {and} \bibinfo{person}{Manish Raghavan}.} \bibinfo{year}{2021}\natexlab{}.
\newblock \showarticletitle{Algorithmic monoculture and social welfare}.
\newblock \bibinfo{journal}{\emph{Proceedings of the National Academy of Sciences}} \bibinfo{volume}{118}, \bibinfo{number}{22} (\bibinfo{year}{2021}), \bibinfo{pages}{e2018340118}.
\newblock


\bibitem[Kornblith et~al\mbox{.}(2019)]%
        {kornblith2019similarity}
\bibfield{author}{\bibinfo{person}{Simon Kornblith}, \bibinfo{person}{Mohammad Norouzi}, \bibinfo{person}{Honglak Lee}, {and} \bibinfo{person}{Geoffrey Hinton}.} \bibinfo{year}{2019}\natexlab{}.
\newblock \showarticletitle{Similarity of neural network representations revisited}. In \bibinfo{booktitle}{\emph{International conference on machine learning}}. PMLR, \bibinfo{pages}{3519--3529}.
\newblock


\bibitem[Lan et~al\mbox{.}(2024)]%
        {lan2024sparse}
\bibfield{author}{\bibinfo{person}{Michael Lan}, \bibinfo{person}{Philip Torr}, \bibinfo{person}{Austin Meek}, \bibinfo{person}{Ashkan Khakzar}, \bibinfo{person}{David Krueger}, {and} \bibinfo{person}{Fazl Barez}.} \bibinfo{year}{2024}\natexlab{}.
\newblock \showarticletitle{Sparse autoencoders reveal universal feature spaces across large language models}.
\newblock \bibinfo{journal}{\emph{arXiv preprint arXiv:2410.06981}} (\bibinfo{year}{2024}).
\newblock


\bibitem[Lenc and Vedaldi(2015)]%
        {lenc2015understanding}
\bibfield{author}{\bibinfo{person}{Karel Lenc} {and} \bibinfo{person}{Andrea Vedaldi}.} \bibinfo{year}{2015}\natexlab{}.
\newblock \showarticletitle{Understanding image representations by measuring their equivariance and equivalence}. In \bibinfo{booktitle}{\emph{Proc. of CVPR}}.
\newblock


\bibitem[Liang et~al\mbox{.}(2020)]%
        {Liang_Li_Li_Wang_Zhang_2020}
\bibfield{author}{\bibinfo{person}{Ruofan Liang}, \bibinfo{person}{Tianlin Li}, \bibinfo{person}{Longfei Li}, \bibinfo{person}{Jing Wang}, {and} \bibinfo{person}{Quanshi Zhang}.} \bibinfo{year}{2020}\natexlab{}.
\newblock \showarticletitle{Knowledge Consistency between Neural Networks and Beyond}.
\newblock  \bibinfo{number}{arXiv:1908.01581} (\bibinfo{year}{2020}).
\newblock
\urldef\tempurl%
\url{http://arxiv.org/abs/1908.01581}
\showURL{%
\tempurl}


\bibitem[Medieros et~al\mbox{.}({[n.\,d.]})]%
        {medieroshuman}
\bibfield{author}{\bibinfo{person}{Kelsey Medieros}, \bibinfo{person}{David~H Cropley}, \bibinfo{person}{Rebecca~L Marrone}, {and} \bibinfo{person}{Roni Reiter-Palmon}.} \bibinfo{year}{[n.\,d.]}\natexlab{}.
\newblock \showarticletitle{Human-AI Co-Creativity: Does ChatGPT make us more creative?}
\newblock  (\bibinfo{year}{[n.\,d.]}).
\newblock


\bibitem[Mednick(1962)]%
        {mednick1962associative}
\bibfield{author}{\bibinfo{person}{Sarnoff Mednick}.} \bibinfo{year}{1962}\natexlab{}.
\newblock \showarticletitle{The associative basis of the creative process.}
\newblock \bibinfo{journal}{\emph{Psychological review}} (\bibinfo{year}{1962}).
\newblock


\bibitem[Moon et~al\mbox{.}(2024)]%
        {moon2024homogenizing}
\bibfield{author}{\bibinfo{person}{Kibum Moon}, \bibinfo{person}{Adam Green}, {and} \bibinfo{person}{Kostadin Kushlev}.} \bibinfo{year}{2024}\natexlab{}.
\newblock \showarticletitle{Homogenizing Effect of Large Language Model (LLM) on Creative Diversity: An Empirical Comparison of Human and ChatGPT Writing}.
\newblock  (\bibinfo{year}{2024}).
\newblock


\bibitem[Olson et~al\mbox{.}(2021)]%
        {olson2021naming}
\bibfield{author}{\bibinfo{person}{Jay~A Olson}, \bibinfo{person}{Johnny Nahas}, \bibinfo{person}{Denis Chmoulevitch}, \bibinfo{person}{Simon~J Cropper}, {and} \bibinfo{person}{Margaret~E Webb}.} \bibinfo{year}{2021}\natexlab{}.
\newblock \showarticletitle{Naming unrelated words predicts creativity}.
\newblock \bibinfo{journal}{\emph{Proceedings of the National Academy of Sciences}} \bibinfo{volume}{118}, \bibinfo{number}{25} (\bibinfo{year}{2021}), \bibinfo{pages}{e2022340118}.
\newblock


\bibitem[Pandya(2024)]%
        {adobe_survey}
\bibfield{author}{\bibinfo{person}{Vivek Pandya}.} \bibinfo{year}{2024}\natexlab{}.
\newblock \showarticletitle{The Age of Generative AI: Over half of Americans have used generative AI and most believe it will help them be more creative}.
\newblock \bibinfo{journal}{\emph{Adobe}} (\bibinfo{year}{2024}).
\newblock
\newblock
\shownote{\url{https://blog.adobe.com/en/publish/2024/04/22/age-generative-ai-over-half-americans-have-used-generative-ai-most-believe-will-help-them-be-more-creative}}.


\bibitem[Papernot et~al\mbox{.}(2016)]%
        {papernot2016transferability}
\bibfield{author}{\bibinfo{person}{Nicolas Papernot}, \bibinfo{person}{Patrick McDaniel}, {and} \bibinfo{person}{Ian Goodfellow}.} \bibinfo{year}{2016}\natexlab{}.
\newblock \showarticletitle{Transferability in machine learning: from phenomena to black-box attacks using adversarial samples}.
\newblock \bibinfo{journal}{\emph{arXiv preprint arXiv:1605.07277}} (\bibinfo{year}{2016}).
\newblock


\bibitem[Pennington et~al\mbox{.}(2014)]%
        {pennington2014glove}
\bibfield{author}{\bibinfo{person}{Jeffrey Pennington}, \bibinfo{person}{Richard Socher}, {and} \bibinfo{person}{Christopher~D. Manning}.} \bibinfo{year}{2014}\natexlab{}.
\newblock \showarticletitle{GloVe: Global Vectors for Word Representation}. In \bibinfo{booktitle}{\emph{Empirical Methods in Natural Language Processing (EMNLP)}}. \bibinfo{pages}{1532--1543}.
\newblock
\urldef\tempurl%
\url{http://www.aclweb.org/anthology/D14-1162}
\showURL{%
\tempurl}


\bibitem[Reimers and Gurevych(2019)]%
        {reimers-2019-sentence-bert}
\bibfield{author}{\bibinfo{person}{Nils Reimers} {and} \bibinfo{person}{Iryna Gurevych}.} \bibinfo{year}{2019}\natexlab{}.
\newblock \showarticletitle{Sentence-BERT: Sentence Embeddings using Siamese BERT-Networks}. In \bibinfo{booktitle}{\emph{Proc. of EMNLP}}. \bibinfo{publisher}{Association for Computational Linguistics}.
\newblock
\urldef\tempurl%
\url{http://arxiv.org/abs/1908.10084}
\showURL{%
\tempurl}


\bibitem[Si et~al\mbox{.}(2024)]%
        {si2024can}
\bibfield{author}{\bibinfo{person}{Chenglei Si}, \bibinfo{person}{Diyi Yang}, {and} \bibinfo{person}{Tatsunori Hashimoto}.} \bibinfo{year}{2024}\natexlab{}.
\newblock \showarticletitle{Can llms generate novel research ideas? a large-scale human study with 100+ nlp researchers}.
\newblock \bibinfo{journal}{\emph{arXiv preprint arXiv:2409.04109}} (\bibinfo{year}{2024}).
\newblock


\bibitem[Stella et~al\mbox{.}(2023)]%
        {stella2023using}
\bibfield{author}{\bibinfo{person}{Massimo Stella}, \bibinfo{person}{Thomas~T Hills}, {and} \bibinfo{person}{Yoed~N Kenett}.} \bibinfo{year}{2023}\natexlab{}.
\newblock \showarticletitle{Using cognitive psychology to understand GPT-like models needs to extend beyond human biases}.
\newblock \bibinfo{journal}{\emph{Proceedings of the National Academy of Sciences}} \bibinfo{volume}{120}, \bibinfo{number}{43} (\bibinfo{year}{2023}), \bibinfo{pages}{e2312911120}.
\newblock


\bibitem[Stevenson et~al\mbox{.}(2022)]%
        {stevenson2022putting}
\bibfield{author}{\bibinfo{person}{Claire Stevenson}, \bibinfo{person}{Iris Smal}, \bibinfo{person}{Matthijs Baas}, \bibinfo{person}{Raoul Grasman}, {and} \bibinfo{person}{Han van~der Maas}.} \bibinfo{year}{2022}\natexlab{}.
\newblock \showarticletitle{Putting GPT-3's creativity to the (alternative uses) test}.
\newblock \bibinfo{journal}{\emph{arXiv preprint arXiv:2206.08932}} (\bibinfo{year}{2022}).
\newblock


\bibitem[Sucholutsky et~al\mbox{.}(2023)]%
        {sucholutsky2023getting}
\bibfield{author}{\bibinfo{person}{Ilia Sucholutsky}, \bibinfo{person}{Lukas Muttenthaler}, \bibinfo{person}{Adrian Weller}, \bibinfo{person}{Andi Peng}, \bibinfo{person}{Andreea Bobu}, \bibinfo{person}{Been Kim}, \bibinfo{person}{Bradley~C Love}, \bibinfo{person}{Erin Grant}, \bibinfo{person}{Iris Groen}, \bibinfo{person}{Jascha Achterberg}, {et~al\mbox{.}}} \bibinfo{year}{2023}\natexlab{}.
\newblock \showarticletitle{Getting aligned on representational alignment}.
\newblock \bibinfo{journal}{\emph{arXiv preprint arXiv:2310.13018}} (\bibinfo{year}{2023}).
\newblock


\bibitem[Szegedy(2014)]%
        {szegedy2013intriguing}
\bibfield{author}{\bibinfo{person}{C Szegedy}.} \bibinfo{year}{2014}\natexlab{}.
\newblock \showarticletitle{Intriguing properties of neural networks}.
\newblock \bibinfo{journal}{\emph{Proc. of ICLR}} (\bibinfo{year}{2014}).
\newblock


\bibitem[Team et~al\mbox{.}(2024a)]%
        {team2024gemini}
\bibfield{author}{\bibinfo{person}{Gemini Team}, \bibinfo{person}{Petko Georgiev}, \bibinfo{person}{Ving~Ian Lei}, \bibinfo{person}{Ryan Burnell}, \bibinfo{person}{Libin Bai}, \bibinfo{person}{Anmol Gulati}, \bibinfo{person}{Garrett Tanzer}, \bibinfo{person}{Damien Vincent}, \bibinfo{person}{Zhufeng Pan}, \bibinfo{person}{Shibo Wang}, {et~al\mbox{.}}} \bibinfo{year}{2024}\natexlab{a}.
\newblock \showarticletitle{Gemini 1.5: Unlocking multimodal understanding across millions of tokens of context}.
\newblock \bibinfo{journal}{\emph{arXiv preprint arXiv:2403.05530}} (\bibinfo{year}{2024}).
\newblock


\bibitem[Team et~al\mbox{.}(2024b)]%
        {team2024jamba}
\bibfield{author}{\bibinfo{person}{Jamba Team}, \bibinfo{person}{Barak Lenz}, \bibinfo{person}{Alan Arazi}, \bibinfo{person}{Amir Bergman}, \bibinfo{person}{Avshalom Manevich}, \bibinfo{person}{Barak Peleg}, \bibinfo{person}{Ben Aviram}, \bibinfo{person}{Chen Almagor}, \bibinfo{person}{Clara Fridman}, \bibinfo{person}{Dan Padnos}, {et~al\mbox{.}}} \bibinfo{year}{2024}\natexlab{b}.
\newblock \showarticletitle{Jamba-1.5: Hybrid Transformer-Mamba Models at Scale}.
\newblock \bibinfo{journal}{\emph{arXiv preprint arXiv:2408.12570}} (\bibinfo{year}{2024}).
\newblock


\bibitem[Van~der Maaten and Hinton(2008)]%
        {van2008visualizing}
\bibfield{author}{\bibinfo{person}{Laurens Van~der Maaten} {and} \bibinfo{person}{Geoffrey Hinton}.} \bibinfo{year}{2008}\natexlab{}.
\newblock \showarticletitle{Visualizing data using t-SNE.}
\newblock \bibinfo{journal}{\emph{Journal of machine learning research}} \bibinfo{volume}{9}, \bibinfo{number}{11} (\bibinfo{year}{2008}).
\newblock


\bibitem[Wang et~al\mbox{.}(2018)]%
        {wang2018great}
\bibfield{author}{\bibinfo{person}{Bolun Wang}, \bibinfo{person}{Yuanshun Yao}, \bibinfo{person}{Bimal Viswanath}, \bibinfo{person}{Haitao Zheng}, {and} \bibinfo{person}{Ben~Y Zhao}.} \bibinfo{year}{2018}\natexlab{}.
\newblock \showarticletitle{With great training comes great vulnerability: Practical attacks against transfer learning}. In \bibinfo{booktitle}{\emph{27th USENIX security symposium (USENIX Security 18)}}. \bibinfo{pages}{1281--1297}.
\newblock


\bibitem[Wu et~al\mbox{.}(2024)]%
        {wu2024generative}
\bibfield{author}{\bibinfo{person}{Fan Wu}, \bibinfo{person}{Emily Black}, {and} \bibinfo{person}{Varun Chandrasekaran}.} \bibinfo{year}{2024}\natexlab{}.
\newblock \showarticletitle{Generative monoculture in large language models}.
\newblock \bibinfo{journal}{\emph{arXiv preprint arXiv:2407.02209}} (\bibinfo{year}{2024}).
\newblock


\bibitem[Zhang et~al\mbox{.}(2024)]%
        {zhang2024generative}
\bibfield{author}{\bibinfo{person}{Simone Zhang}, \bibinfo{person}{Janet Xu}, {and} \bibinfo{person}{A Alvero}.} \bibinfo{year}{2024}\natexlab{}.
\newblock \showarticletitle{Generative ai meets open-ended survey responses: Participant use of ai and homogenization}.
\newblock  (\bibinfo{year}{2024}).
\newblock


\bibitem[Zhao et~al\mbox{.}(2024)]%
        {zhao2024assessing}
\bibfield{author}{\bibinfo{person}{Yunpu Zhao}, \bibinfo{person}{Rui Zhang}, \bibinfo{person}{Wenyi Li}, \bibinfo{person}{Di Huang}, \bibinfo{person}{Jiaming Guo}, \bibinfo{person}{Shaohui Peng}, \bibinfo{person}{Yifan Hao}, \bibinfo{person}{Yuanbo Wen}, \bibinfo{person}{Xing Hu}, \bibinfo{person}{Zidong Du}, {et~al\mbox{.}}} \bibinfo{year}{2024}\natexlab{}.
\newblock \showarticletitle{Assessing and understanding creativity in large language models}.
\newblock \bibinfo{journal}{\emph{arXiv preprint arXiv:2401.12491}} (\bibinfo{year}{2024}).
\newblock


\bibitem[Zhou and Lee(2024)]%
        {zhou2024generative}
\bibfield{author}{\bibinfo{person}{Eric Zhou} {and} \bibinfo{person}{Dokyun Lee}.} \bibinfo{year}{2024}\natexlab{}.
\newblock \showarticletitle{Generative artificial intelligence, human creativity, and art}.
\newblock \bibinfo{journal}{\emph{PNAS nexus}} \bibinfo{volume}{3}, \bibinfo{number}{3} (\bibinfo{year}{2024}), \bibinfo{pages}{pgae052}.
\newblock


\bibitem[Zou et~al\mbox{.}(2023)]%
        {zou2023universal}
\bibfield{author}{\bibinfo{person}{Andy Zou}, \bibinfo{person}{Zifan Wang}, \bibinfo{person}{Nicholas Carlini}, \bibinfo{person}{Milad Nasr}, \bibinfo{person}{J~Zico Kolter}, {and} \bibinfo{person}{Matt Fredrikson}.} \bibinfo{year}{2023}\natexlab{}.
\newblock \showarticletitle{Universal and transferable adversarial attacks on aligned language models}.
\newblock \bibinfo{journal}{\emph{arXiv preprint arXiv:2307.15043}} (\bibinfo{year}{2023}).
\newblock


\end{thebibliography}

\appendix

\vspace{-0.3cm}
\section{Divergent Thinking Test Wording}
\label{appx:dat_prompts}

Here, we report the exact wording for the tests given to humans and LLMs. The wording differs slightly between the two groups because the LLM models are prompted to output their work in a particular format for easier processing, while human prompts refer to text boxes in the survey UI. Without formatting instructions in the prompt, LLMs often discussed the reasoning behind their word choices. While mildly interesting, this muddied the data.

\vspace{-0.3cm}
\subsection{AUT prompts.}
For original experiments, we use the following start words for AUT: \texttt{WORD} = \{book, fork, table, hammer, pants\}. For the expanded LLM evaluation of \S\ref{subsec:family}, we use \texttt{WORD}  = \{book, bottle, brick, fork, hammer, pants, shoe, shovel, table, tire\}.

\para{Human prompt.} {\em Imagine that someone gives you \texttt{WORD}. In the blanks below, write down as many creative uses you can think of for this object, up to 10 uses. }

\para{LLM prompt.}
{\em Imagine that someone gives you a \texttt{WORD}. Write down as many uses as you can think of for this object, up to 10 uses. Please list the uses as words or phrases (single word answers are ok), separated by semicolons. Do not write anything besides your proposed uses.}

\vspace{-0.1cm}
\subsection{Forward Flow prompts.}
We use the following start words for Forward Flow: \texttt{WORD =} \{candle, table, bear, snow, toaster\}. 

\para{Human prompt.} (From the original Flow paper)
{\em Starting with the word \texttt{WORD}, in each of the following blanks, write down the next word that follows in your mind from the previous word. Please put down only single words, and do not use proper nouns (such as names, brands, etc.). Start by writing \texttt{WORD} in the first space below.}

\para{LLM prompt.} {\em Starting with the word \texttt{WORD}, your job is to write down the next word that follows in your mind from the previous word. Please put down only single words, and do not use proper nouns (such as names, brands, etc.). Stop after you listed at least 22 words. Print just the list of words, separated by commas, and do not add anything else to your response. The first word in the list should be 'candle'.}

\vspace{-0.1cm}
\subsection{DAT Prompts.}
\para{Human prompt.} (From the original DAT paper) {\em In the spaces below, please enter 10 words that are as different from each other as possible, in all meanings and uses of the words. You must follow the following rules:
1. Only single words in English.
2. Only nouns (e.g., things, objects, concepts).
3. No proper nouns (e.g., no specific people or places).
4. No specialised vocabulary (e.g., no technical terms).
5. Think of the words on your own (e.g., do not just look at objects in your surroundings).
6. Complete this task in less than four minutes.}

\para{LLM prompt.} {\em Instructions: Please enter 10 words that are as different from each other as possible, in all meanings and uses of the words. Rules: 1. Only single words in English. 2. Only nouns (e.g., things, objects, concepts). 3. No proper nouns (e.g., no specific people or places). 4. No specialised vocabulary (e.g., no technical terms). 5. Think of the words on your own (e.g., do not just look at objects in your surroundings). 6. Complete this task in less than four minutes. 7. Return just the list of words, separated by commas, and do not include any other content.}
\begin{figure*}[h]
    \centering
    \includegraphics[width=0.35\linewidth]{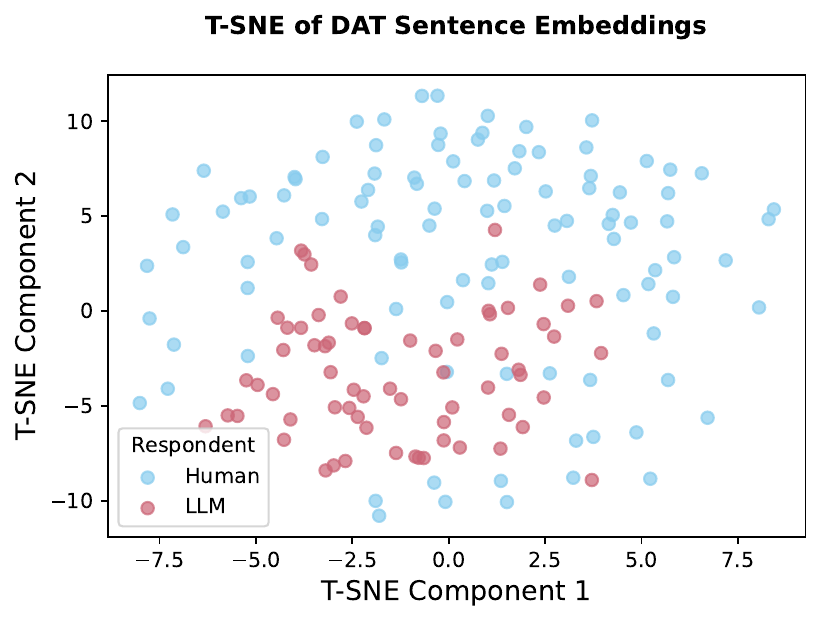}
    \includegraphics[width=0.4\linewidth]{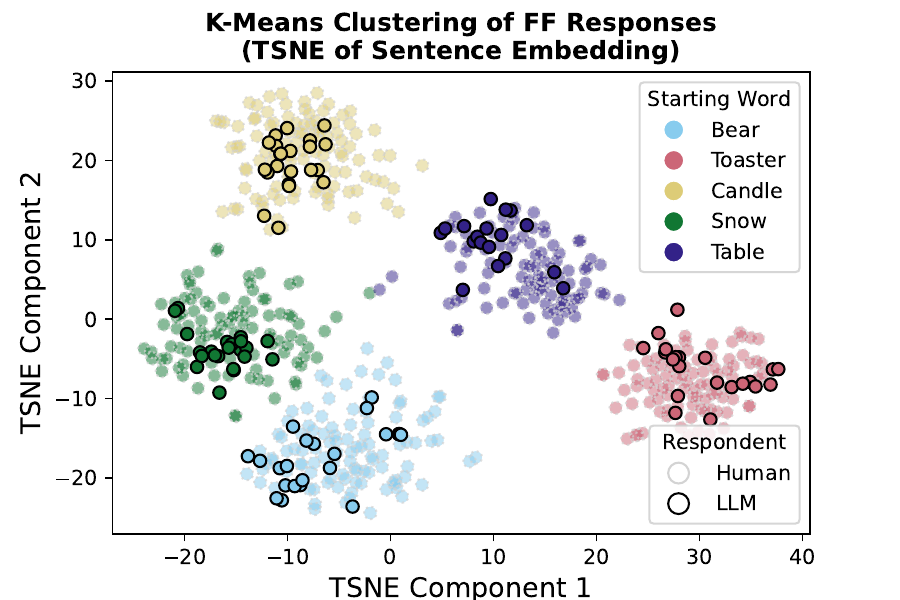}
    \vspace{-0.1cm}
    \caption{{\bf LLM responses to the DAT and FF cluster more in feature space than do human responses}. }
    \label{fig:ff_tsne_dat}
    \vspace{-0.3cm}
\end{figure*}

\section{Originality Scores for AUT, FF, and DAT}
\label{appx:originality}
Here, we describe our methods of computing originality scores for each test. Originality scores are denoted as \orig{t}$(\mathcal{P})$, where $t$ = AUT, FF, or DAT and $\mathcal{P}$ is a population, either humans or LLMs. 

We denote a single word test response as $r$ and an $n$-word test response as ${\bf r} = \{r_0, r_1, \ldots r_n\}$.
The word embedding model is $\mathcal{W}$, and the embedding of a response $r$ is $\mathcal{W}(r)$ (similar for ${\bf r}$, ${\bf r_j}$, etc.). We use cosine similarity $\cos(\mathcal{W}(r_1), \mathcal{W}(r_2))$ to measure semantic distance between embedded responses.
  
\para{AUT scoring.} Following~\cite{dumas2021measuring}, we score the originality of AUT responses by measuring the semantic distance between a prompt $p$ (e.g. ``book'') and each word in $\bf r$ (e.g. ``use it as a doorstop''). Because different words in the AUT response contribute differently to overall response creativity (e.g. ``it'' matters less than ``doorstop''), the final originality score is computed via a weighted sum of these distances. Weights are determined by running TF-IDF analysis on the corpus of responses, which produces low weights for common words like "it" and high weights for unusual words like ``doorstop''. The set of originality scores for AUT responses of population $\mathcal{P}$ is then: 
\begin{equation}
% \origm{AUT}({\mathbf r}, p) = 1 - \frac{\sum_{j=0}^{n-1} w_j \cdot \cos(\mathcal{W}(p), \mathcal{W}(r_j))}{\sum_{j=0}^{n-1} w_j}
\origm{AUT}(\mathcal{P}) = \left\{ 1 - \frac{\sum_{j=0}^{n-1} w_j \cdot \cos(\mathcal{W}(p), \mathcal{W}(r_j))}{\sum_{j=0}^{n-1} w_j}, \forall ({\mathbf r}, p) \in \mathcal{P} \right\}
\end{equation}
where $w_j$ is the TF-IDF weight for the $j^{th}$ word of response $\bf r$ and $p$ is the prompt. %\todo{do we measure similarity or difference} 

\para{ FF scoring}: Here, we follow the methodology of~\cite{gray2019forward}. This defines the ``instantaneous'' forward flow of a given thought in the sequence $\bf r$ as the average distance between the $m^{th}$ thought in the sequence $r_m$ and all preceding thoughts:
$$
\frac{\sum_{j=1}^{m-1} (1-\cos(\mathcal{W}(r_j), \mathcal{W}(r_{m})))}{m-1}
$$ Building on this, the set of FF scores for a population $\mathcal{P}$ consisting of $n$-word sequences $\bf r$ is given by: \begin{equation}
\origm{FF}(\mathcal{P}) = \left\{ \frac{1}{n-1} \cdot  \sum_{i=2}^{n} \frac{\sum_{j=1}^{i-1} (1-\cos(\mathcal{W}(r_j), \mathcal{W}(r_{i})))}{(i-1)}, \forall ~{\mathbf r} \in \mathcal{P} \right\}
\end{equation}

\para{DAT scoring:} We use the scoring methodology of~\cite{olson2021naming}, which scores responses by averaging the semantic distance between all pairs of words in the response. Given a population $\mathcal{P}$ composed of $n$-long DAT response $\bf r$ containing words $\{r_0, r_1, \ldots r_{n-1}\}$, the set of DAT scores is calculated as:
\begin{equation}
\origm{DAT}(\mathcal{P}) = \left\{ \frac{1}{n(n-1)} \sum_{i,j (i \ne j)}^{n-1} (1 - \cos(\mathcal{W}(r_i), \mathcal{W}(r_j))) \forall ~\mathbf{r} \in \mathcal{P} \right\}
\end{equation}

\section{TSNE of FF and DAT}
\label{appx:tsne_others}

Figure~\ref{fig:ff_tsne_dat} visualizes the TSNE of sentence embeddings for DAT and FF responses, similar to Figure~\ref{fig:aut_tsne_sent}. This confirms the trend observed in the AUT TSNE: LLM responses cluster closer in feature space than human responses, resulting in lower population-level originality measurements. We perform k-means clustering of TNSE of FF sentence embeddings to demonstrate clusters of LLM response for each start word. Since the DAT since the test does not involve varying start words, we simply visualize all LLM and human responses.

\end{document}